\documentclass{emulateapj}
\pdfoutput=1

\shorttitle{Multi-D Double Detonation of White Dwarfs}
\shortauthors{Moll \& Woosley}

\usepackage{amsmath}
\usepackage{natbib}
\usepackage{xspace}
\usepackage{multirow}
\usepackage[percent]{overpic}
\usepackage[colorlinks=true,linkcolor=blue,citecolor=blue]{hyperref}

\newcommand{\Msun}{\ensuremath{M_\odot}}
\newcommand{\unit}[1]{\ensuremath{\,\mathrm{#1}}}
\newcommand{\gcc}{\ensuremath{\unit{g\;cm^{-3}}}}
\newcommand{\cms}{\ensuremath{\unit{cm\;s^{-1}}}}
\newcommand{\degree}{\ensuremath{^\circ}}
\newcommand{\simm}{\ensuremath{\mathord{\sim}}}
\newcommand{\el}[2]{$^{#2}\text{#1}$}
\newcommand{\mel}[2]{^{#2}\text{#1}}
\newcommand{\de}{\ensuremath{\mathrm{d}}}

\newcounter{mod}
\newcounter{sim}

\newcommand{\myscale}{1}

\begin{document}

\title{Multi-Dimensional Double Detonation of Sub-Chandrasekhar Mass White Dwarfs}

\author{R.~Moll and S.~E.~Woosley}

\affil{Department of Astronomy and Astrophysics, University of California, Santa Cruz, CA 95064, USA}
 
\date{\today}

\begin{abstract}
Using 2D and 3D simulation, we study the ``robustness'' of the double
detonation scenario for Type Ia supernovae, in which a detonation in
the helium shell of a carbon-oxygen white dwarf induces a secondary
detonation in the underlying core.  We find that a helium detonation
cannot easily descend into the core unless it commences (artificially)
well above the hottest layer calculated for the helium shell in
current presupernova models.  Compressional waves induced by the
sliding helium detonation, however, robustly generate hot spots which
trigger a detonation in the core. Our simulations show that this is
true even for non-axisymmetric initial conditions. If the helium is
ignited at multiple points, the internal waves can pass through one
another or be reflected, but this added complexity does not defeat
the generation of the hot spot.  The ignition of very low-mass helium
shells depends on whether a thermonuclear runaway can simultaneously
commence in a sufficiently large region.
\end{abstract}


\section{Introduction}

In the double detonation scenario, a carbon-oxygen (CO) white dwarf (WD)
accumulates a layer of helium through accretion from a helium main sequence
star.  The accretion rate, typically on the order of $10^{-8}\text{--}10^{-7}
\Msun\unit{yr^{-1}}$ \citep[e.g.,][]{1980Taam,1994Woosley}, and the mass of the
WD determine the critical thickness of the helium layer. Generally, a low
accretion rate allows more helium to accumulate before a thermonuclear runaway
commences near the base of the helium shell, the temperature of which is
determined by gravitational compression, convection and eventually nuclear
burning.  It is often assumed that a detonation in the helium induces a
secondary detonation in the CO core, thus producing a Type Ia supernova.

If and how a secondary core detonation develops is ambiguous, however.  One
possibility is that the core is ignited directly at the core-shell interface
when hit by the helium detonation front. This is sometimes called ``direct
drive'', ``edge-lit scenario'' or ``prompt detonation''. It has long been
realized that this does not always work and that its realization depends on the
altitude at which the helium detonation starts
\citep[e.g.,][]{1990Livne,1997Benz,1999Garcia}.  A helium detonation directly
at the interface almost certainly fails to start a detonation in the core, but
evolution models of accreting WDs suggest that the most probable place for a
detonation to start could be higher up \citep{2011Woosley}. One of the aims of
the present work is to determine whether a helium detonation ignited at a point
in the hottest helium layer would lead to core detonation by direct drive.

If direct drive fails, the core may still ignite, with considerable
delay, after the detonation wave has consumed the entire helium layer,
through compressional waves which focus inside the core.  In the
simplest (and most unrealistic) scenario, the star is spherically
symmetric and the helium is ignited instantaneously in a spherical
shell. This induces a radially inward propagating compressional wave
that converges at the exact center of the star, where, depending upon
the resolution of the study, it always ignites the core
\cite[e.g.,][]{1990aLivne}.

Since convective mixing is not able to keep the temperature at a given
radius constant before the runaway, it is more likely that helium
ignition starts in one or more isolated regions with sizes of
perhaps a pressure scale height \citep{2011Woosley}.  If the
detonation starts at a single point, the sliding helium detonation
creates an oblique (with respect to the core-shell interface) shock
wave that converges inside the core at the axis of symmetry and
potentially triggers an off-center core detonation
\citep{1990Dgani,1990Livne,1991Livne,1995Livne}.  In their simulations
of double-detonation supernovae, \citet{2007Fink} considered a variety
of 1D and 2D helium ignition scenarios, confirming the high ignition
potential of the inward propagating shock wave.  They showed that
core detonation resulting from the detonation of the helium in a
single spot is very likely even for low mass helium shells, for instance
$0.0035\Msun$ in combination with a $1.3850\Msun$ core
\citep{2010Fink}, and for core masses as low as $0.45\Msun$
\citep{2012Sim}.

In a scenario with less or no symmetry, if helium detonations commence
at multiple points, for example, the secondary detonation of the core
is less certain.  In their SPH simulations, \citet{1999Garcia} first
explored, using coarse resolution, the possibility of starting
detonations at multiple points in the helium layer. In particular,
they studied a case with asynchronous ignition at five points, finding
that core detonation could still result from the collision of the various
helium detonations.

If the core does not ignite, the helium explosion alone could produce a
sub-luminous, ``point Ia'', supernova \citep{2007Bildsten}.  This raises the
question as to how robust the triggering of the secondary core detonation is.
If it is robust, a detonation of the helium shell alone might be very rare or
even impossible in the case of CO cores (the less common oxygen-neon cores are
in general much harder to ignite with converging shocks, see
\citealt{2013Shen}).

We here revisit the double detonation scenario using some of the WD models
generated with KEPLER and presented in \citet{2011Woosley}.  Following a
description of the numerical methods in \S\ref{sec:methods}, we first discuss
the results of simulations that test the viability of direct drive in
\S\ref{sec:directdrive}.  We then (\S\ref{sec:hotspots}) study sliding helium
detonations that lead to a hot spot inside the core, with the helium detonation
started at one, two or three points.  The initiation of detonations in
lightweight helium shells is investigated in \S\ref{sec:lighthesh}.  In
\S\ref{sec:complete}, we present nucleosynthesis yields from runs following the
complete detonation of WDs. We end with a discussion and summary in
\S\ref{sec:discussion}.

\section{Numerical Methods}
\label{sec:methods}

\begin{deluxetable*}{lccccccccc}
\tablecaption{White dwarf models}
\tablehead{
model\tablenotemark{a}
         & $M_\text{core}$  & $M_\text{He}$      & $r_\text{core}$
         & $\Delta r_\text{He}$\tablenotemark{b} & $r(T_\text{max})$      & $T_\text{max}$
         & $\rho\bigl(r(T_\text{max})\bigr)$  &  $\rho(r_\text{core})$  & $\rho(r=0)$ \\
         &  [$\Msun$]       & [$\Msun$]          & [km]
         &  [km]                  & [km]  & [$10^8\unit{K}$]
         &  [$10^6\gcc$]    & [$10^6\gcc$]       & [$10^7\gcc$]
}
\startdata
\refstepcounter{mod}\label{mod:p8b1a}
\Alph{mod} (8B) & 0.801 & 0.143 & 4081 & 2080 & 4267 & 2.06 & 1.19 & 1.66 & 2.21   \\   
\refstepcounter{mod}\label{mod:1pa1a}
\Alph{mod} (10B)   & 1.000 & 0.082 & 3616 & 1370 & 3765 & 2.42 & 1.30 & 1.93 & 5.67   \\   
\refstepcounter{mod}\label{mod:1pb1a}
\Alph{mod} (10C)   & 1.000 & 0.091 & 3518 & 1380 & 3624 & 2.31 & 1.63 & 2.18 & 6.02   \\
\refstepcounter{mod}\label{mod:1pi1a}
\Alph{mod} (10HC)  & 1.002 & 0.045 & 4161 & 1260 & 4187 & 2.80 & 0.717 & 0.932 & 4.24   \\
\refstepcounter{mod}\label{mod:1pj1a}
\Alph{mod} (10HD)  & 1.001 & 0.078 & 3696 & 1350 & 3724 & 2.36 & 1.50 & 1.77 & 5.33
\enddata
\tablenotetext{a}{The model identifiers in \citet{2011Woosley} are written in parentheses}
\tablenotetext{b}{Approximate helium shell thickness, with the outer boundary
defined where the density is $1/100$ of the value at the core-shell interface}
\label{tab:models}
\end{deluxetable*}

\subsection{Hydrodynamics}
\label{sec:hydro}

For our multi-dimensional studies we use the Eulerian hydrodynamics code CASTRO
\citep{2010Almgren,2011Zhang} to solve the equations for compressible fluid
dynamics in combination with a nuclear reaction network.  We employ an equation
of state based on the Helmholtz free energy \citep{1999Timmes,2000Timmes} and a
19-isotope network to compute energy generation and nucleosynthesis (the same
as in KEPLER, see \S\ref{sec:nucegen} for details).  Gravitational forces are
calculated with a monopole approximation, taking into account the mass enclosed
within a given distance from the center of the star at any given time.  In many
simulations, we make use of CASTRO's capabilities for adaptive mesh refinement
(AMR).  The initial models for the multi-D CASTRO simulations presented here
were constructed with the implicit Lagrangian hydrodynamics code KEPLER in 1D
\citep{1978Weaver,2002Woosley}.  The evolution of several helium-accreting WDs
was followed with a model for convection based on mixing-length theory, see
\citet{2011Woosley}.  The CASTRO simulations start from KEPLER snapshots taken
before a thermonuclear runaway sets in.  Their key properties are listed in
Table~\ref{tab:models}.

In Eulerian simulations, adding a relatively small amount of hot gas to a cold
zone can raise the average temperature above the threshold for ignition, thus
igniting the whole zone.  Without any correction, the speed with which a
``detonation front'' propagates would thus be dependent on the size of the
zones and driven entirely by artificial advection.  To simulate detonation more
realistically, we introduce a delay timer that suppresses the burning in a
particular zone for half a sound crossing time. If, at any time, the necessary
conditions for nuclear burning are not satisfied, the timer is set to zero.  If
they are satisfied, the timer starts counting and when half a sound crossing
time (Courant time) has passed, nuclear burning processes are switched on in
this particular zone.

In simulations of the whole star, the density of the artificial external medium
is set to $1\gcc$ and given the same temperature as the outer edge of the
helium shell.  This means that as the shock of the helium detonation breaks out
of the star, it is accelerated by a very steep pressure gradient. As the time
step of our simulations is limited by the largest fluid velocities or sound
speeds in the entire domain, the calculations become very expensive. To contain
the detrimental effect on the time step, we apply a velocity cap of
$1.5\times10^9\cms$ in the ambient medium. To distinguish the ambient from
stellar material at all times of the simulation, we give it a unique
composition of 100\% \el{O}{16}.

We used no velocity cap and lower ambient densities in runs where the expansion
of the ejecta was followed (\S\ref{sec:complete}).  Whenever the ejecta reached
the boundaries of the current computational domain, the solution was embedded
into a new, larger domain (twice as large in each dimension, but with the same
resolution at the coarsest level). The embedded solution was initially kept at
a higher mesh refinement level (such that it expands into the coarser grid,
rather than being mapped into it).  This step was repeated several times until
homologous expansion was reached. The final domain had an extent of
$1.6\times10^9\unit{cm}$ in cylindrical radius and twice that value in the
$z$-direction.  During the remaps into larger domains, the densities of the
ambient media were lowered by factors of 10 from $10^{-2}\gcc$ (in the original
domain) to $10^{-5}\gcc$. While ambient material piles up at the explosion
front, the total mass of this material is low, on the order of $10^{-4}\Msun$,
and we do not expect it to have significant influence on either the ejecta
structure or the velocities.

We generally make use of symmetries to reduce the computational load.  With
only one detonator, the problem is axisymmetric and its solution can be
computed in 2D.  For a simulation with two synchronous spherical detonators it
is sufficient to do a 3D calculation of only a quarter of the star, with mirror
symmetry at two boundaries (the $x=0$ and $y=0$ boundaries in our simulations).
Two asynchronous detonators require a half-star simulation with mirror symmetry
at one boundary (the $x=0$ boundary in our simulations).

\subsection{Nuclear Energy Generation and Nucleosynthesis}
\label{sec:nucegen}

\begin{figure}[t]
\centering
\includegraphics[width=.8\linewidth]{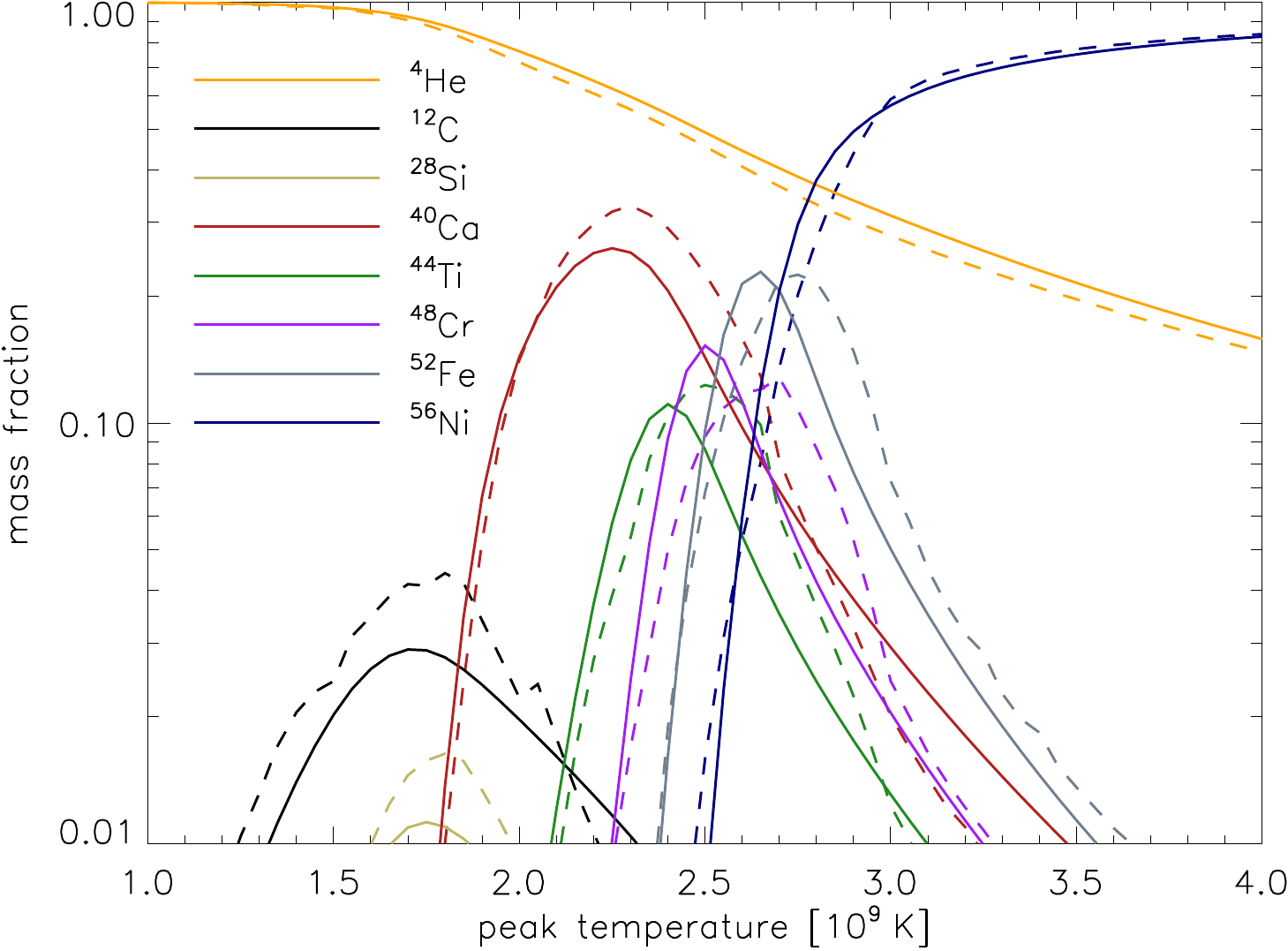}
\caption{Variation of the final composition in explosive helium burning,
expandend along paths parameterized by the peak temperature $T_0$ (see text).
The solid and dashed lines represent the results obtained with the full network
and the table, respectively.}
\label{fig:hetabtest}
\end{figure}

Energy generation and approximate nucleosynthesis were calculated in all cases
using a 19-isotope network with constituent species as defined by
\citet{1978Weaver} and nuclear reaction rates updated to current values
\citep{2007Woosley}. Screening corrections were implemented for all rates. This
network includes an ``alpha-chain'' ($Z = N = 2n$, $n = \text{an integer}$) of
nuclei from \el{C}{12} through \el{Ni}{56}, plus the species n, p, \el{He}{3},
\el{N}{14}, and \el{Fe}{54}. Simulated steady state links (i.e., $\de Y_i / \de
t \approx 0$ within a time step) also approximate the presence of the odd $Z$
isotopes from \el{Na}{23} through \el{Co}{55} (i.e., isotopes with $Z = 2n+1$,
$N = 2n+2$).  For this study, the photodisintegration of \el{Ni}{56} to
$\mel{Fe}{54} + 2\mathrm{p}$ was not allowed which meant that electron capture
and \el{Fe}{54} were essentially omitted from the network. Furthermore
\el{H}{1}, \el{He}{3}, \el{N}{14}, n, and p never achieved any significant
abundances, so the effective network really just contained the 13 alpha-chain
isotopes.  Past studies \citep{2000bTimmes} have shown that such a limited
network gives good agreement with the results of much larger networks provided
one is interested only in bulk nucleosynthesis and energy generation.

The network itself was employed in all carbon-rich zones always and, in most
cases, for the helium shell. In some cases, however, it was computationally
expedient to use a table that was prepared off-line using the network. Those
cases included studies of helium shell detonation where the whole star was
carried (\S\ref{sec:hotspots}), up until the generation of a hot spot in the
core.  The simulations for which detailed yields are given
(\S\ref{sec:complete}) make use of the full network from beginning to end, as
do the local simulations of direct drive (\S\ref{sec:directdrive}).

The use of a table here was novel and warrants some discussion. Unlike isobaric
burning in e.g. carbon deflagration supernovae, the table here has to work
reliably in dynamic situations where both the temperature and density are
rapidly varying and where the outcome is sensitive to the time history of both.
Using the network, a large table was generated using the results of many
off-line studies of helium burning at constant temperature and density. Values
of temperature in the range $10^{8.5}\text{--}10^{9.65}\unit{K}$ ($\Delta \log
T = 0.05$) and density in the range $10^5\text{--}10^{7.1}\gcc$ ($\Delta \log
\rho = 0.10$) were included. Results of the burning---the composition, average
binding energy per nucleon, and time derivative of the helium mass
fraction---were sampled for 50 values of helium mass fraction ranging from 0.02
to 1.0 in steps of 0.02. To preserve accuracy, the rate equation for the helium
mass fraction was divided by the leading order term in the $3 \alpha$ reaction
rate, i.e.,
\begin{equation}
\left(\frac{\de X_\text{He}}{\de t}\right)_\text{table} \ = \ \left(\frac{\de
  X_\text{He}}{\de t}\right)_\text{calc} T_9^3 \left(\rho^2
X_{\alpha}^3 \exp\left(\frac{-4.403}{T_9}\right) \right)^{-1}.
\end{equation}
This resulted in a slowly varying function of $T_9$, $\rho$, and
$X_{\alpha}$ in the table.

When using the table in a multi-dimensional simulation, triple variable
interpolation in $\log\rho$, $\log T_9$, and $X_\text{He}$ gave the current
composition and $\de X_\text{He}/\de t$. The current time step and $\de
X_\text{He}/\de t$ gave a new helium abundance at the end of the step. Another
call to the table using the new helium abundance gave a revised composition for
the other species, and the change in composition gave a change in nuclear
binding energy, hence an energy generation, and the process continued.

The shortcoming of this approach is that the entries in the table were
generated at constant temperature. This neglects additional variables because a
given temperature, density, and helium abundance can be reached by a variety of
histories and the instantaneous abundances of the other nuclei are sensitive to
which history was followed. Some additional rules are imposed to avoid
unphysical and poorly behaved solutions. The helium abundance was only allowed
to decrease, i.e., $\de X_\text{He}/\de t$ had to be negative, or no
composition change or energy generation is allowed. Further, no modification of
the composition occurred unless both the abundance of \el{Ni}{56} and the nuclear
binding energy increased during a time step. When these conditions were met, a
change in composition was allowed, but the new composition was not just taken
from the table. Instead a fraction of the new composition at the given
temperature and density was added to the old one with that fraction determined
by the change of the helium abundance during the step. Essentially this
procedure partitions the nuclear evolution for an arbitrarily varying
temperature and density history into a series of small steps taken at constant
temperature and density. For the conditions of the problem, the requirement
that the time step be a small fraction of the Courant time proved adequate to
preserve accuracy.

To demonstrate the validity of the procedure, a series of simulated explosive
helium burning studies were carried out off line using both the network and the
table.  Starting with a composition of pure helium, matter was expanded from a
peak temperature $T_0$ and a corresponding peak density
\begin{equation}
    \rho_0 = 3 \left(\frac{T_0}{3.5\times10^9\unit{K}}\right)^2 \times 10^6 \gcc
\end{equation}
along paths given by
\begin{align}
                       T &= T_0 \exp(-t/\tau) \\
    \text{and}\quad \rho &= \rho_0 \exp(-3t/\tau)
\end{align}
with $\tau=0.2\unit{s}$ being the time scale for the explosion.  The final
composition was examined when the temperature had declined below
$10^8\unit{K}$.  The results are shown in Figure~\ref{fig:hetabtest}.  In
general the agreement between the network results and those obtained using the
table is quite good.  Using the table gives a smaller final abundance for
helium because burning at a constant low temperature gives a smaller mean
nuclear mass and charge than explosive burning cooling from a higher
temperature to the same helium mass fraction, density and temperature. Lighter
nuclei have larger cross sections for $\alpha$-capture. One could improve the
fit by adding a fourth independent variable, the mean nuclear charge (excluding
helium itself), but the present approach was deemed adequate for this study.
Some test runs at constant temperature and density using the table and the
network gave near perfect agreement showing that the table had been correctly
implemented.

We further tested the tables by calculating the yields of 2D helium detonations
in model~\ref{mod:p8b1a}, with a setup similar to run~\ref{sim:p0406x}
presented in \S\ref{sec:results}, but using closed boundaries at all sides to
prevent hot ash from leaving the computational domain (which we wanted to keep
small for this test).  The detonation wave converges after $\mathord{\approx}
1.20\unit{s}$ on the other side of the star. Compared to the calculation with
the table, the detonation with the network is only slightly faster (the
detonation front is ahead by about $4\degree$ as measured from the center at
$t=1.15\unit{s}$, with about half of the discrepancy being due to a slower
start of the artificial detonator with the table).  We measured the yields at
$t=1.70\unit{s}$, at which time the helium burning has ceased (we suppressed
the burning of carbon). With the network, we thus get a \el{Ni}{56} yield of
$0.0721\Msun$, burning $0.0934\Msun$ of \el{He}{4}. With the table, the numbers
are $0.0687$ and $0.0909$. The energy produced is $2.78\times10^{50}\unit{erg}$
with the network, and $2.66\times10^{50}\unit{erg}$ with the table. This is
adequate for our purposes.

\section{Results}
\label{sec:results}

\subsection{Direct Drive}
\label{sec:directdrive}

Here we consider the possibility of a helium detonation wave that traverses the
core-shell interface beneath its origin, directly igniting the core at the
outer edge. In the first two sections, we present several simulations
calculated in 2D using CASTRO. Pursuing the hypothesis that direct drive is
related to a critical mass, we then discuss the results of idealized 1D
simulations using the KEPLER code in \S\ref{sec:ddrivekepler}.

\renewcommand{\myscale}{.25}
\begin{figure*}[t]
\centering
\begin{tabular}{llll}
(a) &
(b) &
(c) &
(d) \\
\includegraphics[scale=\myscale]{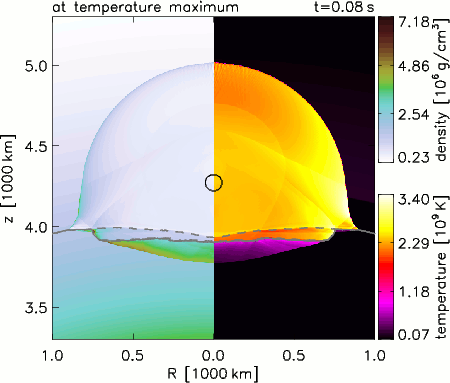} &
\includegraphics[scale=\myscale]{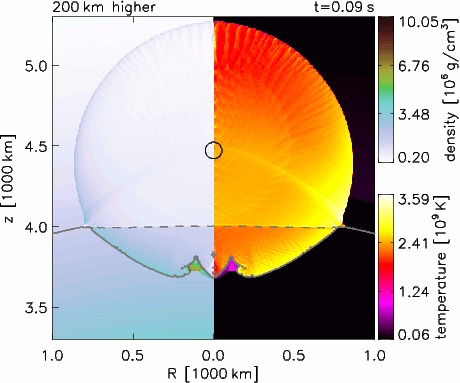} &
\includegraphics[scale=\myscale]{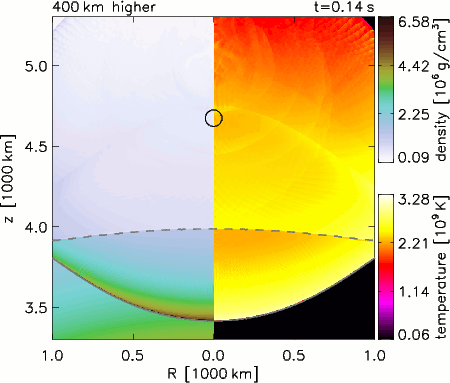} &
\includegraphics[scale=\myscale]{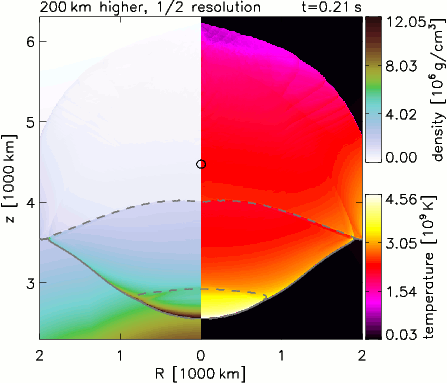} \\
(e) &
(f) &
(g) &
(h) \\
\includegraphics[scale=\myscale]{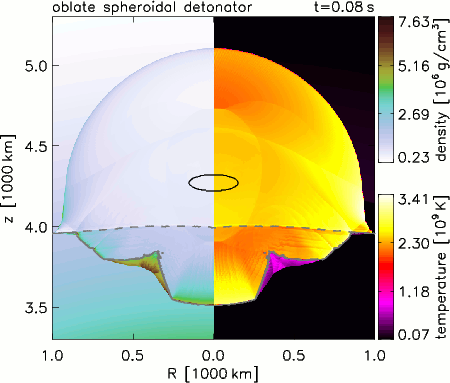} &
\includegraphics[scale=\myscale]{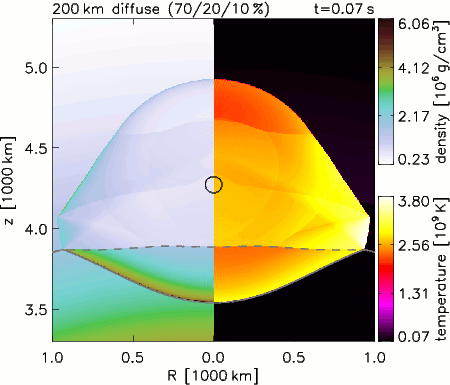} &
\includegraphics[scale=\myscale]{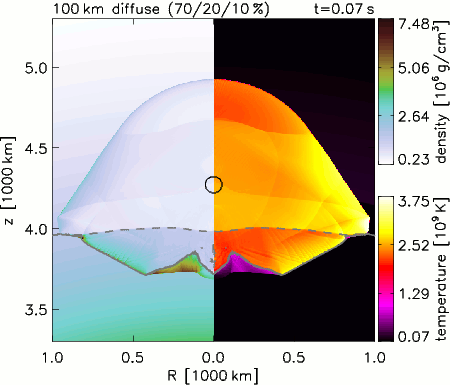} &
\includegraphics[scale=\myscale]{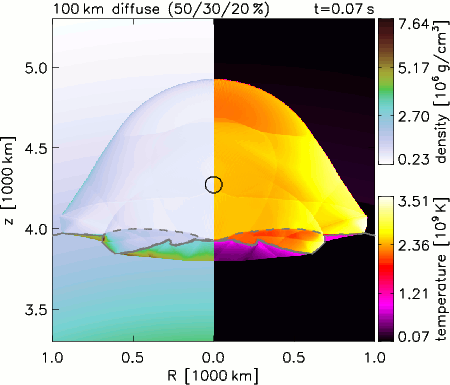}
\end{tabular}
\caption{Density (left, upper colorbar) and temperature (right, lower colorbar)
in local 2D simulations of helium detonations hitting upon the carbon-oxygen
core.  The black circles indicate the perimeters of the helium detonation
spots at the beginning of each simulation.  The solid and dashed gray lines
represent the mass fraction contours for 49\% \el{C}{12} and 49\% \el{O}{16},
respectively.}
\label{fig:dd}
\end{figure*}

\begin{figure}[t]
\centering
\includegraphics[width=.7\linewidth]{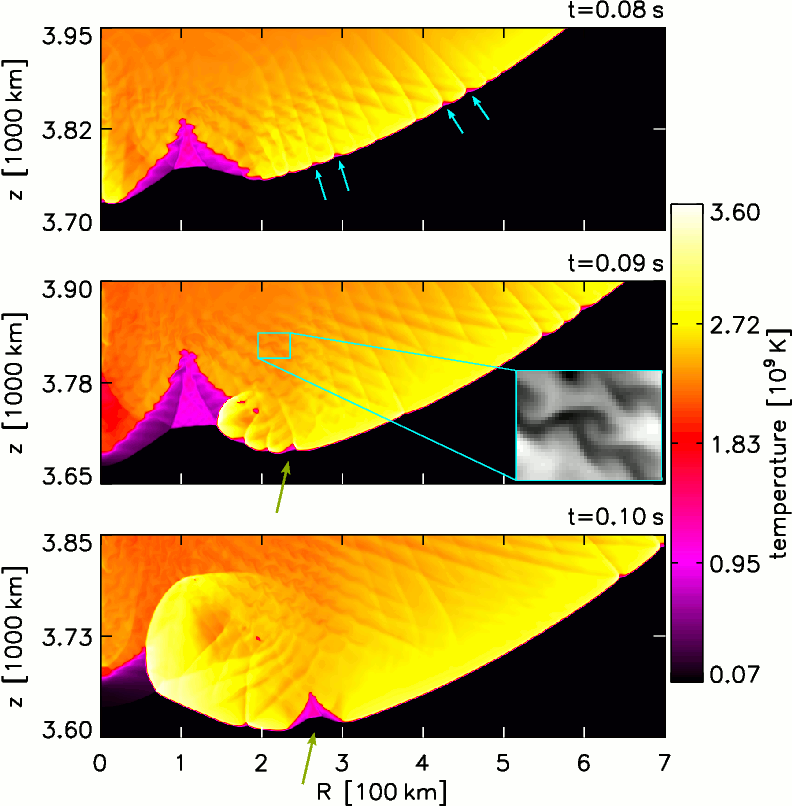}
\caption{Snapshots of the detonation front in case (b) of the models for direct
drive discussed in \S\ref{sec:ddrivexpl}. The panel in the middle is a blow-up
of Figure~\ref{fig:dd}b; the inset zooms in on vortical structures in the ash,
with the temperatures in the plot ranging from
$(2.24\ldots2.45)\times10^9\unit{K}$ (black to white).  The arrows indicate
indentations in the detonation front.}
\label{fig:cusp}
\end{figure}

\begin{deluxetable}{ccccc}
\tablecaption{Direct drive in model~\ref{mod:p8b1a}}
\tablehead{
\multirow{2}{*}{run\tablenotemark{a}}  & $\Delta x$  & altitude\tablenotemark{b} & \multirow{2}{*}{specialty\tablenotemark{c}}  & \multirow{2}{*}{works/fails} \\
                        &  [km]               & [km]       &       & 
}
\startdata
a & 0.977 & 0 & --- & fails \\
b & 0.977 & 200 & --- & fails \\
c & 0.977 & 400 & --- & works \\
d & 1.953 & 200 & --- & works \\
e & 0.977 & 0 & 3:1 spheroidal detonator & fails \\
-- & 0.977 & 0 & 4:1 spheroidal detonator & fails \\
f & 0.977 & 0 & \raggedright mixed (70\% \el{He}{4}) a\&b & works \\
g & 0.977 & 0 & \raggedright mixed (70\% \el{He}{4}) & fails \\
-- & 1.953 & 0 & \raggedright mixed (70\% \el{He}{4}) & fails \\
h & 0.977 & 0 & \raggedright mixed (50\% \el{He}{4}) & fails \\
-- & 0.977 & 0 & \raggedright mixed (80\% \el{He}{4}) & fails
\enddata
\tablenotetext{a}{Labels as in Figure~\ref{fig:dd}; ``--'' denotes runs mentioned in the text only}
\tablenotetext{b}{Distance to the hottest helium layer}
\tablenotetext{c}{``mixed'' stands for a mixed layer of helium and core material above and below (``a\&b'')
or only below the original core-shell interface}
\label{tab:ddriva}
\end{deluxetable}

\begin{deluxetable}{ccccc}
\tablecaption{Direct drive: masses of the helium detonation spheres at core contact}
\tablehead{
\multirow{2}{*}{model}  & $r_\text{detonator}$  & $\Delta r$\tablenotemark{a} & $M_\text{He-sphere}$\tablenotemark{b}  & \multirow{2}{*}{works/fails} \\
                        &  [km]               & [km]       & [g]       & 
}
\startdata
\ref{mod:p8b1a} & 4270 & 189 & $3.89\times10^{28}$ & fails \\
\ref{mod:p8b1a} & 4470 & 389 & $2.80\times10^{29}$ & fails \\
\ref{mod:p8b1a} & 4670 & 589 & $8.28\times10^{29}$ & works \\
\ref{mod:1pa1a} & 3765 & 149 & $2.17\times10^{28}$ & fails \\
\ref{mod:1pa1a} & 3815 & 199 & $4.75\times10^{28}$ & fails \\
\ref{mod:1pa1a} & 3865 & 249 & $8.71\times10^{28}$ & works \\
\ref{mod:1pb1a} & 3624 & 106 & $9.40\times10^{27}$ & fails \\
\ref{mod:1pb1a} & 3636 & 118 & $1.26\times10^{28}$ & fails \\
\ref{mod:1pb1a} & 3649 & 131 & $1.68\times10^{28}$ & works \\
\ref{mod:1pb1a} & 3674 & 156 & $2.75\times10^{28}$ & works \\
\ref{mod:1pb1a} & 3724 & 206 & $5.91\times10^{28}$ & works
\enddata
\tablenotetext{a}{Distance from the core-shell interface to the detonator center}
\tablenotetext{b}{Defined as twice the mass of the helium contained in the half-sphere of radius $\Delta r$
below $z=r_\text{detonator}$}
\label{tab:ddrive}
\end{deluxetable}

\subsubsection{Altitude, Shape and Interface Composition}
\label{sec:ddrivexpl}

Taking the $(0.801+0.143)\Msun$ WD model~\ref{mod:p8b1a} as initial condition,
we did a series of 2D simulations that cover a local region including the
core-shell interface.  The domain range in most cases is $0<R_8<1$ and
$3.3<z_8<5.3$ in cylindrical coordinates $(R,z)$, the core-shell interface
being at the spherical radius $r_8=4.08$. The grid resolution in most cases is
$0.977\unit{km}$ (no AMR). The gravitational field is taken to be static and
the boundary conditions are usually reflecting on all sides (as the detonation
proceeds supersonically, the choice of boundary conditions is not critical).
All simulations employed a full 19-isotope network which covers helium, carbon
and oxygen burning. The detonation spots used to start the detonation had a
radius of $50\unit{km}$, a temperature of $2\times10^9\unit{K}$ at the center
and $1.8\times10^9\unit{K}$ at the outer edge (linearly decreasing), a radial
velocity of $8000\unit{km\;s^{-1}}$ at the outer edge (increasing linearly from
zero) and twice the ambient density.

Table~\ref{tab:ddriva} contains a list of setups and results, and
Figure~\ref{fig:dd} depicts snapshots of several cases.  In the standard case
(a), the helium detonator is centered at the radius at which the temperature is
maximal, $186\unit{km}$ above the core-shell interface. While the detonation
creates an indentation in the interface, and some of the carbon at the outer
edge of the core burns as a result of the entering shock, the burning does not
support a detonation in the core. Direct drive clearly fails in this case.

With the helium detonation initiated $200\unit{km}$ higher, case (b), the
detonation penetrates into the core.  The front is unstable though, developing
dents filled with compressed material that fails to ignite.  The formation of
these instabilities appears to be the result of complex ineractions of
reflecting shock waves.  Figure~\ref{fig:cusp} shows snapshots of the
temperature in the detonation front in case (b).  There are small indentations
in the front, seemingly related to waves in the post-detonation region (top
panel, indicated by cyan arrows).  A few hundred kilometers below the
core-shell interface, a big dent forms (at $R \leq 200\unit{km}$ in the plots).
While it is later on consumed by a secondary detonation front, a new dent forms
(middle and bottom panels, olive-green arrows). The vortical structures visible
in the plots on a high zoom level are indicative of Richtmyer-Meshkov
instabilities \citep[RMI, e.g.,][]{2002Brouillette}. The presence of RMI is
supported by the fact that the pressure gradient behind the detonation is
normal to the front, whereas the density gradient across the interface is
radial with respect to the center of the star, i.e., the pressure and density
gradients are misaligned, as is typical for RMI.  It is possible, however, that
RMI here emerge as a secondary phenomenon.  Global simulations, presumably in
3D, would be needed to determine whether such an unstable detonation eventually
blows out or consumes the core, but they are too expensive at this resolution.
Recognizing that these are marginal cases, we opted not to follow such
pathological detonations in this study and dismissed them as failures.

With the helium detonation initiated $400\unit{km}$ above the temperature
maximum, case (c), the detonation stably penetrates the core.  This strengthens
our assumption that the instabilities described above are indicative of
marginal cases.  To make sure that the detonation stays stable, we followed it
in a global simulation that encompasses half of the star, with the same
resolution ($\mathord{\approx}1\unit{km}$) as in the local simulation in a
region that extends from the detonation spot to the center of the star ($R_8<1$
and $z_8<5$). This detonation stably proceeds all the way to the center.
Oxygen burning starts $830\unit{km}$ below the interface (at 80\% of the core
radius).

In a simulation similar to case (b) with half resolution ($1.95\unit{km}$),
case (d), the front is stable.  The formation of indentations in the front is
completely suppressed.  We followed this detonation until a depth of
$1780\unit{km}$ beneath the core-shell interface.  Oxygen starts to burn
$1130\unit{km}$ below the interface (at 72\% of the core radius, see the lower
dashed line in panel d), where the density exceeds $5.3\times10^6\gcc$.

The fronts of detonations started at different altitude have different
curvatures when they hit upon the core-shell interface. This raises the
question as to whether the curvature of the detonation front plays a key role.
Apart from altitude, the curvature is determined by the shape of the detonation
spot, provided the spot is sufficiently close to the interface (for isotropic
expansion velocities, any explosion eventually becomes spherical). A detonator
in the form of an oblate spheroid fares better than a spherical one, compare
panels (e) and (a). However, the detonation front also starts out considerably
hotter than in the spherical case ($T_9^\text{max}=3.24$ instead of $2.88$ at
$t=0.01\unit{s}$), indicating that part of the observed difference is possibly
related to the details of how the detonation was started, rather than curvature
alone\footnote{
The spheroidal detonation spot was constructed using the same temperature and
velocity profiles with respect to the distance of the surface of the spot as
for spherical detonators. However, this means that the energy contents (kinetic
and thermal) are different. The distance to the core-shell interface may not be
large enough for the detonation wave to lose all information of its
(artificial) initiation besides curvature.  This complicates the comparison.
However, as the explosions induced by the spheroidal detonators are in general
stronger (consistent with a higher energy content), we can still draw
conclusions about the non-importance of curvature if direct drive with such a
detonator fails.
}.
The detonation spot in case (e) is stretched by a factor 3 in radial direction,
its extent in the vertical direction being the same as that of the spherical
spots.  Shortly before it reaches the core-shell interface, the curvature of
the underside of the oblate detonation front roughly corresponds to a sphere
centered $200\unit{km}$ above the hottest layer, i.e., it is comparable to the
detonation in case (b).  A simulation with a detonator stretched by a factor 4
(i.e., even more oblate) gives a very similar result (not shown in
Figure~\ref{fig:dd}).  The curvature in this case corresponds to a sphere
centered more than $400\unit{km}$ above the hottest layer, i.e., one would
expect from case (c) that the detonation should enter the core without
problems, yet it does not. We therefore surmise that curvature is not as
important as altitude.

We also considered the possibility of a layer where helium mixes with core
material. In case (f), we start with a mixture of 70/20/10\%
\el{He}{4}/\el{C}{12}/\el{O}{16} in the region $3.98 < r_8 < 4.18$, i.e., $100\unit{km}$
above \textsl{and} below the original core-shell interface. The detonation
stably transcends into the core in this case. We surmise that a mixed
layer on both sides of the interface effectively increases the distance from
the detonator to the core, thus facilitating direct drive.

In cases (g) and (h), we start with a mixture of 70/20/10\%
\el{He}{4}/\el{C}{12}/\el{O}{16} and 50/30/20\%
\el{He}{4}/\el{C}{12}/\el{O}{16}, respectively, in a $100\unit{km}$ wide region
above the original interface only. Although the detonation changes when it
enters the mixed region, becoming hotter and faster, it fails to penetrate the
core and stay intact in both cases.  Even with half resolution, the detonation
in case (g) shows the same instability (unlike case d, for which the coarser
resolution helped to suppress instabilities in the detonation front).  A
mixture of 80/20\% \el{He}{4}/\el{C}{12} in a $100\unit{km}$ wide region also
leads to failure (not shown in Figure~\ref{fig:dd}).

\subsubsection{Heavy White Dwarfs}

The above discussion shows that direct drive is not likely for
model~\ref{mod:p8b1a} if the helium detonation starts where the gas is
hottest, i.e., the most natural location for a
detonator. Models~\ref{mod:1pa1a} and~\ref{mod:1pb1a} both have a
denser 1.0 $\Msun$ core which in principle should detonate more easily
than the $0.8\Msun$ core of model~\ref{mod:p8b1a}.
Model~\ref{mod:1pb1a} is slightly more compact than
model~\ref{mod:1pa1a}.

Following a similar procedure, we tested for the possibility of direct
ignition with detonators at different altitudes for these models. The
radius of the detonation ``hot spots'' was $20\unit{km}$ and the distance
between the core-shell interface and the temperature maximum was
$149\unit{km}$ and $106\unit{km}$ in models~\ref{mod:1pa1a}
and~\ref{mod:1pb1a}, respectively. In both cases, direct ignition
failed if the detonation was started at this altitude.

If we started the helium detonation $100\unit{km}$ above the hottest layer of
model~\ref{mod:1pa1a} however, the detonation front smoothly transcended into
the core.  If ignited $50\unit{km}$ above the hot layer, the detonation
fragmented and died upon passing into the core. In model~\ref{mod:1pb1a}, the
direct drive worked with detonators at altitudes of $100\unit{km}$,
$50\unit{km}$ and even $25\unit{km}$ above the hottest layer. At $12\unit{km}$,
however, it failed\footnote{Given that this displacement is becoming comparable
to the size of the detonator itself, it is questionable whether direct drive
will or will not work for the most massive models. Also the exact location of
of the hot layer is probably not known to this precision due to the
one-dimensional mixing length approximation used to treat convection.
Multi-dimensional simulation of the convection during the pre-explosive runaway
might provide additional insight.}. These results for direct drive with
detonators at different altitudes are summarized in Table~\ref{tab:ddrive}.

\subsubsection{The Critical Altitude for Direct Drive}
\label{sec:ddrivekepler}

To explore the relevant physics underlying the existence of a critical altitude
for directly driving a detonation into the carbon-oxygen core beneath the
ignition point in the helium layer, a series of idealized calculations were
carried out using the KEPLER code.  The hypothesis to be explored was that
successful propagation requires a critical amount of momentum in the downwardly
moving shock wave. That momentum should be greater than or equal to that of a
``critical mass'' \citep{1997Niemeyer,2007Roepke,2009Seitenzahl} of carbon and
oxygen at the same density as exists at the helium-carbon interface. In
addition, that momentum should be focused into an area comparable to the
geometrical size of the critical mass. Stated another way, the altitude of the
initial helium detonation should exceed the radius of the critical mass
required for initiating a successful carbon detonation evaluated at a density
equal to that at the helium-carbon interface.

Of course this is quite approximate. The density and pressure are not constant
in either the helium layer or the carbon underneath, and the energy yield from
helium detonation is different from that of carbon detonation. We do not know
if the helium detonation ignites at a point, which implies spherical
propagation, or in an extended region which might have a more planar geometry.
These complexities can be partly compensated for, however. The effective radius
of the helium is one that encloses a critical \textsl{mass} (evaluated at the
density of the interface), even though the density within that sphere varies.
The detonation can be studied in a two phase medium with a helium detonator
surrounded by a thick shell of carbon and oxygen. The detonation should occur
in spherical geometry so long as the size of the region in the helium layer
that intially runs away is small compared with its altitude above the
interface. We note that if, pending further study, the latter assumption were
violated, the required altitude would be reduced (compare Tables 1 and 4 in
\citet{2009Seitenzahl}).

In the first study, spheres of 50\% carbon, 50\% oxygen were prepared with a
constant density of $1.93 \times 10^6\gcc$, as appropriate to the interface in
model~\ref{mod:1pa1a}. In a one-dimensional spherical region of variable
radius, the temperature was given a constant gradient (with respect to radius,
not mass, even though the code is Lagrangian) with values ranging from $2.8
\times 10^9\unit{K}$ at the center to $1.0 \times 10^8\unit{K}$ at the edge.
This 200 zone region was surrounded by 300 zones of carbon and oxygen at
$10^8\unit{K}$.  All zones had the same radial thickness. The radius of the
region with the temperature gradient was varied and its ability to generate a
self-sustaining detonation determined. The central value of $2.8 \times
10^9\unit{K}$ was sufficiently high to guarantee that carbon burning propagated
with a phase velocity that was initially supersonic.  This procedure is very
similar to the one previously followed by \citet{1997Niemeyer} and
\citet{2007Roepke}.

For a ``detonator'' radius of 150 km, an initially strong detonation decayed
away after traversing an additional 70 km of ``cold'' carbon and oxygen. For a
radius of 200 km however, the detonation propagated successfully to the edge of
the grid at 500 km. This is sufficiently far, considering that in the real star
the density would have become higher and the detonation more robust. These
detonators enclosed masses of $2.73 \times 10^{28}\unit{g}$ ($150\unit{km}$)
and $6.47 \times 10^{28}\unit{g}$ ($200\unit{km}$) respectively. 

A more precise descriptor of the critical detonation size may be the
temperature gradient rather than the critical mass. For the two runs described
here, the temperature gradients are $120\unit{K\;cm^{-1}}$ (failed detonation)
and $90\unit{K\;cm^{-1}}$ (successful) respectively. \citet{2009Seitenzahl}
studied the initiation of carbon detonation in spherical geometry and found a
critical gradient of $4360\unit{K\;cm^{-1}}$ at a density of $5 \times
10^6\gcc$ (their Table 4). Our value is consistent with an admittedly uncertain
extrapolation of their Figure 9 to $1.93 \times 10^6\gcc$.

More realistically though, the detonator consists of helium, not carbon and
oxygen, and the initial background is better approximated as isobaric rather
assuming than constant density.  An initial temperature of $2.8 \times
10^9\unit{K}$ in the carbon also implies an unrealistically high energy
density, not likely to be achieved at these low densities.  A pure helium
sphere with variable radius was surrounded by a thick shell of 50\% \el{C}{12}
and 50\% \el{O}{16} again at $1.93\times10^6\gcc$ with a temperature of $1.0
\times 10^8\unit{K}$. The helium core was comprised of 200 Lagrangian mass
shells of equal mass with a central temperature of only $1.0 \times
10^9\unit{K}$ and a temperature at the edge of $1.0 \times 10^8\unit{K}$ and a
constant gradient (with respect to radius) in between. As with the carbon
detonator, the large value of central temperature assured a phase velocity for
the helium burning that was initially supersonic and the well zoned gradient
implied the existence of a region where it was sonic. The density within the
helium core was varied so as to provide a constant pressure equal to the
pressure in the cold carbon-oxygen layer.  Thus pressure on the entire grid was
initially $7.25 \times 10^{22}\unit{dyne\;cm^{-2}}$. Gravity was neglected and
the pressure was initially held constant by applying a boundary pressure equal
to the pressure on the grid.  If the radius of the helium sphere was
$211\unit{km}$ enclosing $6.37 \times 10^{28}\unit{g}$, the detonation
successfully propagated into the surrounding carbon. Another run with a helium
sphere radius of $159\unit{km}$ ($2.69 \times 10^{28}\unit{g}$) failed to
detonate the carbon shell however.

The energy yield from carbon detonation at this density, taken from the
calculation, is $3.3 \times 10^{17}\unit{erg\;g^{-1}}$. Oxygen does not burn
here, but carbon and neon do, to magnesium and silicon. The yield from helium
detonation, on the other hand, is only $(5\ldots6) \times
10^{17}\unit{erg\;g^{-1}}$ because only about one-third of the helium burns to
\el{Ni}{56} immediately behind the shock (the helium burning reaction rate
saturates at high temperature). Some helium continues to burn well after the
shock has passed, eventually raising the energy yield to $7 \times
10^{17}\unit{erg\;g^{-1}}$, but this energy is not so important for sustaining
the shock wave. The critical radius is also the cube root of the critical mass.
This explains the similarity between the runs using a carbon-oxygen detonator
and helium detonator at this density.

These results agree qualitatively with those obtained using a 2D representation
of the full star in the CASTRO code (Table~\ref{tab:ddrive}).  They also agree
reasonably well with the pure carbon-oxygen study at constant density described
above. Apparently the critical altitude (or detonator radius) is not very
sensitive to the different energy yields of carbon and helium burning. This is
somewhat surprising, but confirms the hypothesis that the critical altitude in
the helium layer is one that would enclose the critical mass of carbon required
for detonation at the interface density.

\subsection{Shock Collisions and Hot Spots}
\label{sec:hotspots}

\begin{deluxetable*}{cccccccccccc}
\tablecaption{Conditions at the hot spot induced by converging compressional waves in the core}
\tablehead{
\multirow{2}{*}{run} & \multirow{2}{*}{model}    & \multirow{2}{*}{$\alpha$\tablenotemark{a}}
       & $\tau$\tablenotemark{b} & $\Delta x$\tablenotemark{c}    & $\Delta t$\tablenotemark{d}  
       & $t$  & $z$              & $r$\tablenotemark{e} & $T$              & $\rho$  & \multirow{2}{*}{type\tablenotemark{f}}     \\
       &      &                      & [s] & [km]                           & [s]
       & [s]  & [$10^8\unit{cm}$]& [$10^8\unit{cm}$]    & [$10^9\unit{K}$] & [$10^7\gcc$] &
}
\startdata
\refstepcounter{sim}\label{sim:p0403x}
\arabic{sim} & \ref{mod:p8b1a}  & --- & --- &            6.51    & 0.05     & 1.50 & $-1.90$    & 1.90           & 5.25      & 6.21  & p   \\   
\refstepcounter{sim}\label{sim:p0422x}
\arabic{sim} & \ref{mod:p8b1a}& 180\degree & ---   & 6.51    & 0.05 & 1.40 & $\pm1.45$   & 1.45   & 1.93     & 3.55 & r  \\
\refstepcounter{sim}\label{sim:p0425x}
\arabic{sim} & \ref{mod:p8b1a}& 180\degree & 0.30  & 6.51    & 0.05 & 1.45 & $-1.72$   & 1.72   & 3.34    & 4.00 & r \\
\refstepcounter{sim}\label{sim:p0229x}
\arabic{sim} & \ref{mod:p8b1a}    & --- & --- &            14.6    & 0.02     & 1.48 & $-1.87$    & 1.87           & 8.09      & 2.68 & p     \\   
                 &     &     & & 3.66\tablenotemark{g}    &  0.02  & 1.48 & $-1.92$    & 1.92           & 9.61      & 2.54 & p     \\   
\refstepcounter{sim}\label{sim:p0406x}
\arabic{sim} & \ref{mod:p8b1a}    & --- & --- &             25.0    & 0.05     & 1.55 & $-1.96$    & 1.96           & 3.09      & 4.29 & p     \\   
                 &     &     &  &           25.0    & 0.005    & 1.515 & $-1.91$   & 1.91           & 5.51      & 20.9  & p   \\   
\refstepcounter{sim}\label{sim:h0407x}
\arabic{sim} & \ref{mod:p8b1a}   & $54\degree$ & --- &      25.0    & 0.05     & 1.40 & $-1.76$    & 1.77        & 2.81      & 5.66  & p     \\
\refstepcounter{sim}\label{sim:h0427x}
\arabic{sim} & \ref{mod:p8b1a}   & $54\degree$ & 0.15 &     25.0    & 0.05     & 1.45 & $-1.74$    & 1.80        & 3.67      & 9.51   & p    \\
\refstepcounter{sim}\label{sim:h0430x}
\arabic{sim} & \ref{mod:p8b1a} & $90\degree$ & --- &   25.0    & 0.05   & 1.40  & $-1.39$    & 1.54        & 2.17      & 5.04   & t    \\
\refstepcounter{sim}\label{sim:j0510x}
\arabic{sim} & \ref{mod:p8b1a} & $120\degree$ & 0.30 &   25.0    & 0.05   & 1.50  & $-0.91$    & 1.96        & 2.86      & 3.59   & t    \\
\refstepcounter{sim}\label{sim:h0529x}
\arabic{sim} & \ref{mod:p8b1a} & $120\degree$, $90\degree$, $100\degree$ & 0.30, 0.20 &   25.0    & 0.05   & 1.45  & $-0.71$    & 1.65  & 2.19  & 5.27  & t   \\
\refstepcounter{sim}\label{sim:p0324x}
\arabic{sim}\tablenotemark{h} & \ref{mod:p8b1a}& --- & --- & 52.1   & 0.05     & 1.55 & $-1.95$    & 1.95           & 2.96    & 7.56   & p    \\
\refstepcounter{sim}\label{sim:p0315x}
\arabic{sim} & \ref{mod:p8b1a}   & $36\degree$ & --- &      52.1    & 0.05     & 1.45 & $-1.74$    & 1.74        & 2.19      & 4.93   & p    \\
\refstepcounter{sim}\label{sim:p0322x}
\arabic{sim} & \ref{mod:p8b1a}   & $54\degree$ & --- &      52.1    & 0.05     & 1.45 & $-1.74$    & 1.81        & 2.00      & 4.24  & p   \\
\refstepcounter{sim}\label{sim:p0514x}
\arabic{sim} & \ref{mod:1pj1a}   & --- & --- &     6.51    & 0.05     & 1.40 & $-1.70$    & 1.40           & 3.52      & 4.54   & p   \\ 
\refstepcounter{sim}\label{sim:p0515x}
\arabic{sim} & \ref{mod:1pj1a}  & $180\degree$ & --- &    6.51    & 0.05     & 0.60 & 0    & 3.48           & 3.48      & 0.60  & b   \\
\refstepcounter{sim}\label{sim:p0627x}
\arabic{sim} & \ref{mod:1pi1a}   & --- & --- &     6.51    & 0.05     & 1.80 & $-2.13$    & 2.13           & 2.44     & 3.31   & p   \\ 
\refstepcounter{sim}\label{sim:p0625x}
\arabic{sim} & \ref{mod:1pi1a}  & $180\degree$ & --- &    6.51    & 0.05     & --- & ---    & ---           & ---      & ---  & --- 
\enddata
\tablenotetext{a}{Angular separation(s) of the detonators in cases with multiple detonators}
\tablenotetext{b}{Delay of the second (and third) detonator}
\tablenotetext{c}{Grid resolution at the highest AMR level that includes the hot spot}
\tablenotetext{d}{Cadence of the considered snapshots}
\tablenotetext{e}{Distance from the center (spherical radius)}
\tablenotetext{f}{Type of listed hot spot: (p) where primary waves converge, (r) where reflected waves collide,
(t) where a wave passes through others and converges or (b) near the core boundary, where primary waves first collide}
\tablenotetext{g}{Restarted with finer zoning in the region where the internal wave converges}
\tablenotetext{h}{Despite this being a 2D problem, this run was done in 3D}
\label{tab:hspots}
\end{deluxetable*}

\renewcommand{\myscale}{.27}
\begin{figure*}[t]
\center
\begin{tabular}{lll}
(a) run~\ref{sim:p0406x}, one detonator &
(b) run~\ref{sim:h0407x}, two detonators, $54\degree$ &
(c) run~\ref{sim:h0430x}, two detonators, $90\degree$ \\
\includegraphics[scale=\myscale]{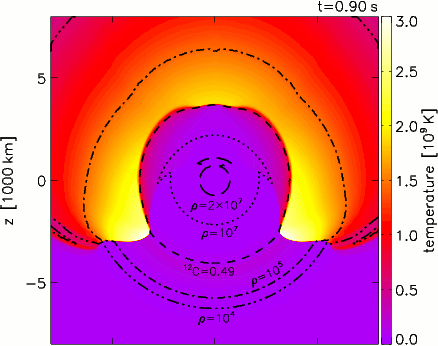} &
\includegraphics[scale=\myscale]{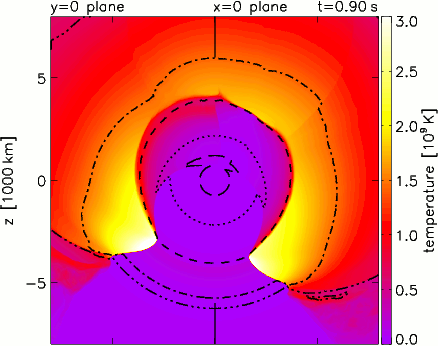} &
\includegraphics[scale=\myscale]{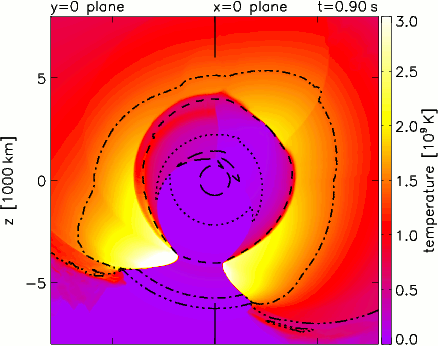} \\
\includegraphics[scale=\myscale]{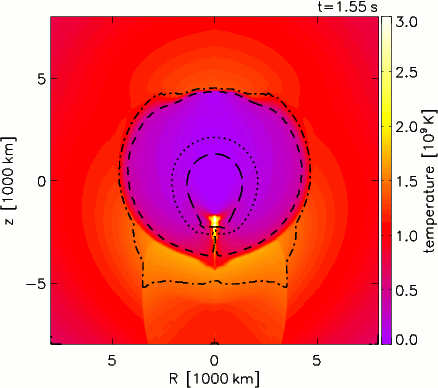} &
\includegraphics[scale=\myscale]{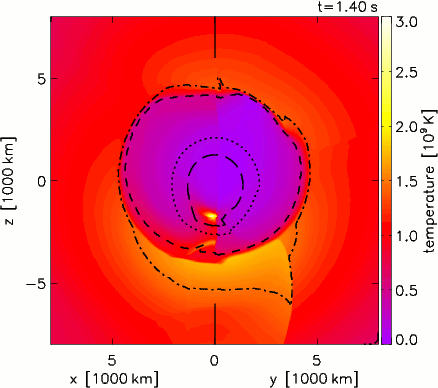} &
\includegraphics[scale=\myscale]{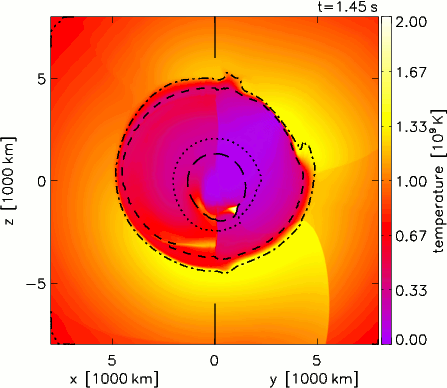}
\end{tabular}
\caption{Snapshots of simulations with one (a) and two synchronous detonators
(b: $54\degree$ separation, c: $90\degree$ separation) during the helium
detonation stage (upper panels) and during the presence of a hot spot in the
core (lower panels). Panels (b) and (c): the $x=0$ plane (right half), which
contains the two detonators and the center of the star, is symmetric with
respect to the $y=0$ plane (left half) and vice versa; the long ticks at $R=0$
indicate that these plots show not one but two orthogonal planes.  The black
contour lines represent, from the center outwards: $\rho=2\times10^7\gcc$ (long
dashed), $\rho=10^7\gcc$ (dotted), $X_\mathrm{C}=0.49$ (dashed),
$\rho=10^5\gcc$ (dot-dashed) and $\rho=10^4\gcc$ (dot-dot-dot-dashed).}
\label{fig:hotspot}
\end{figure*}

\renewcommand{\myscale}{.26}
\begin{figure*}[t]
\centering
\begin{tabular}{ccc}
\includegraphics[scale=\myscale]{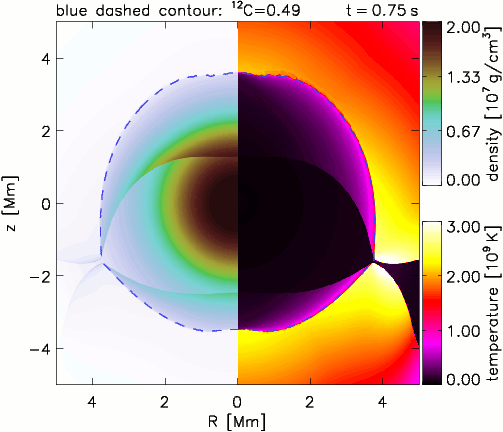}
\includegraphics[scale=\myscale]{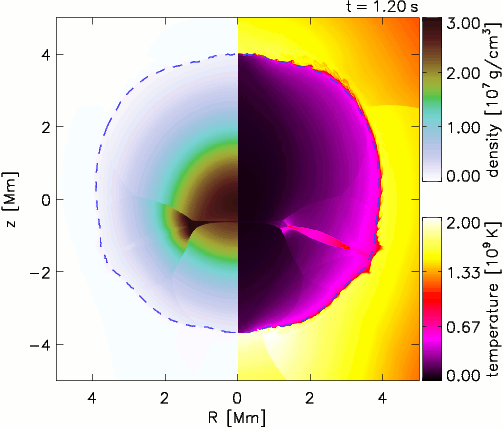}
\includegraphics[scale=\myscale]{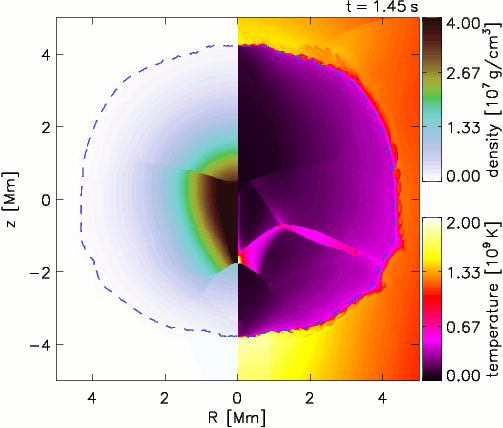}
\end{tabular}
\caption{Three snapshots of a 2D simulation with asynchronous detonators on
opposite sides (run~\ref{sim:p0425x}): when the helium detonation fronts
collide (left), when the internal wave converges (middle) and when reflected
waves produce a hot spot (right). The blue dashed line indicates the core-shell
boundary. Only the inner part of the simulation box is shown.}
\label{fig:p0425}
\end{figure*}

\begin{figure*}[t]
\centering
\begin{overpic}[width=.89\linewidth]{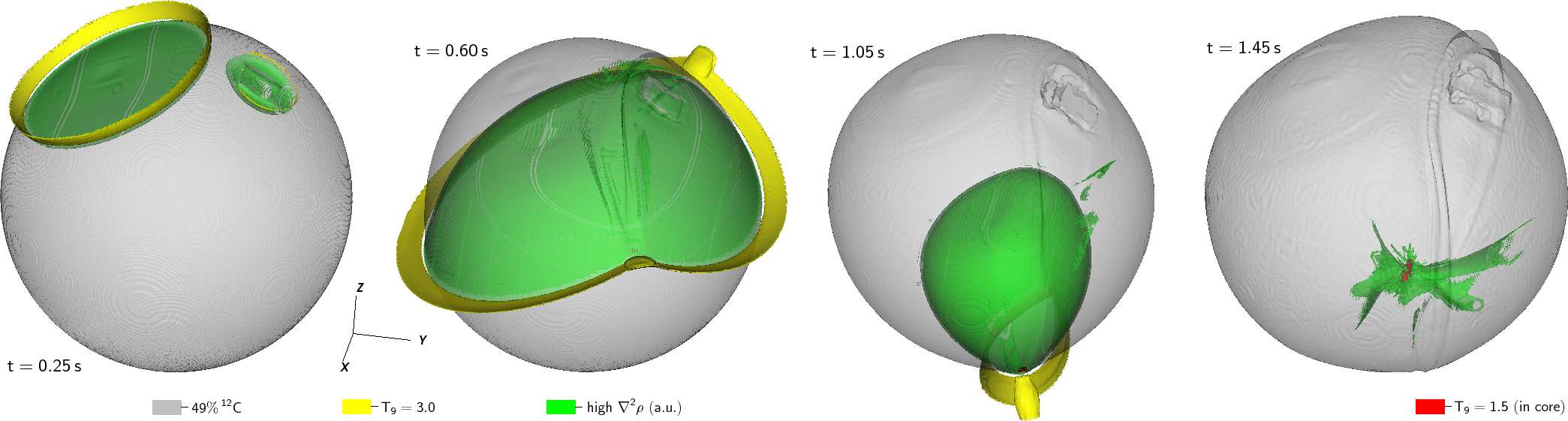}
\put(70,0){\includegraphics[width=.08\linewidth,trim=20 0 0 5,clip]{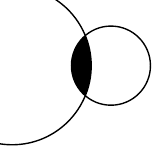}}
\end{overpic}
\caption{Helium shell detonation started by two asynchronous detonators with a
$54\degree$ separation in the hottest helium layer (run~\ref{sim:h0427x}). The
gray, yellow and green isosurfaces indicate the boundary of the CO core, the
detonation front and the internal compressional wave, respectively. The sketch
next to the snapshot at $t=1.05\unit{s}$ (second from right) qualitatively
mimics the shape of the converging detonation front as the intersection of two
circles. Red isosurfaces indicate the hot spot that is generated when the
internal wave converges (rightmost snapshot).}
\label{fig:h0427threed}
\end{figure*}

\begin{figure}[t]
\center
\begin{tabular}{c}
\includegraphics[width=.8\linewidth]{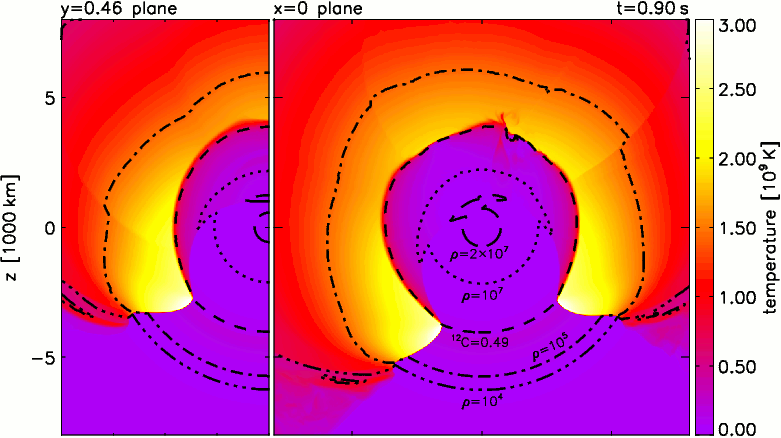} \\
\includegraphics[width=.8\linewidth]{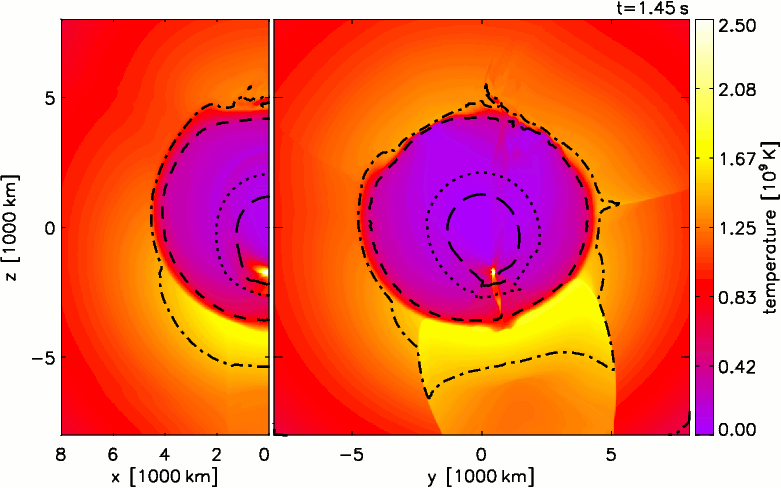}
\end{tabular}
\caption{Snapshots of the temperature in a simulation with two asynchronous
detonators at $\pm27\degree$ from the positive $z$-axis on the $x=0$ plane
(run~\ref{sim:h0427x}). The $x=0$ plane (right panels) contains the detonators
and the center of the star, and the orthogonal $y_8=0.46$ plane (left panels)
contains a hot spot generated by converging internal waves. The detonator on
the positive side of the $y$-axis is delayed by $0.15\unit{s}$.  The black
contour lines represent, from the center outwards: $\rho=2\times10^7\gcc$ (long
dashed), $\rho=10^7\gcc$ (dotted), $X_\mathrm{C}=0.49$ (dashed),
$\rho=10^5\gcc$ (dot-dashed) and $\rho=10^4\gcc$ (dot-dot-dot-dashed). The
snapshot at the bottom is taken when the internal wave fronts converge.}
\label{fig:h0427x}
\end{figure}

\begin{figure}[t]
\center
\includegraphics[width=.8\linewidth]{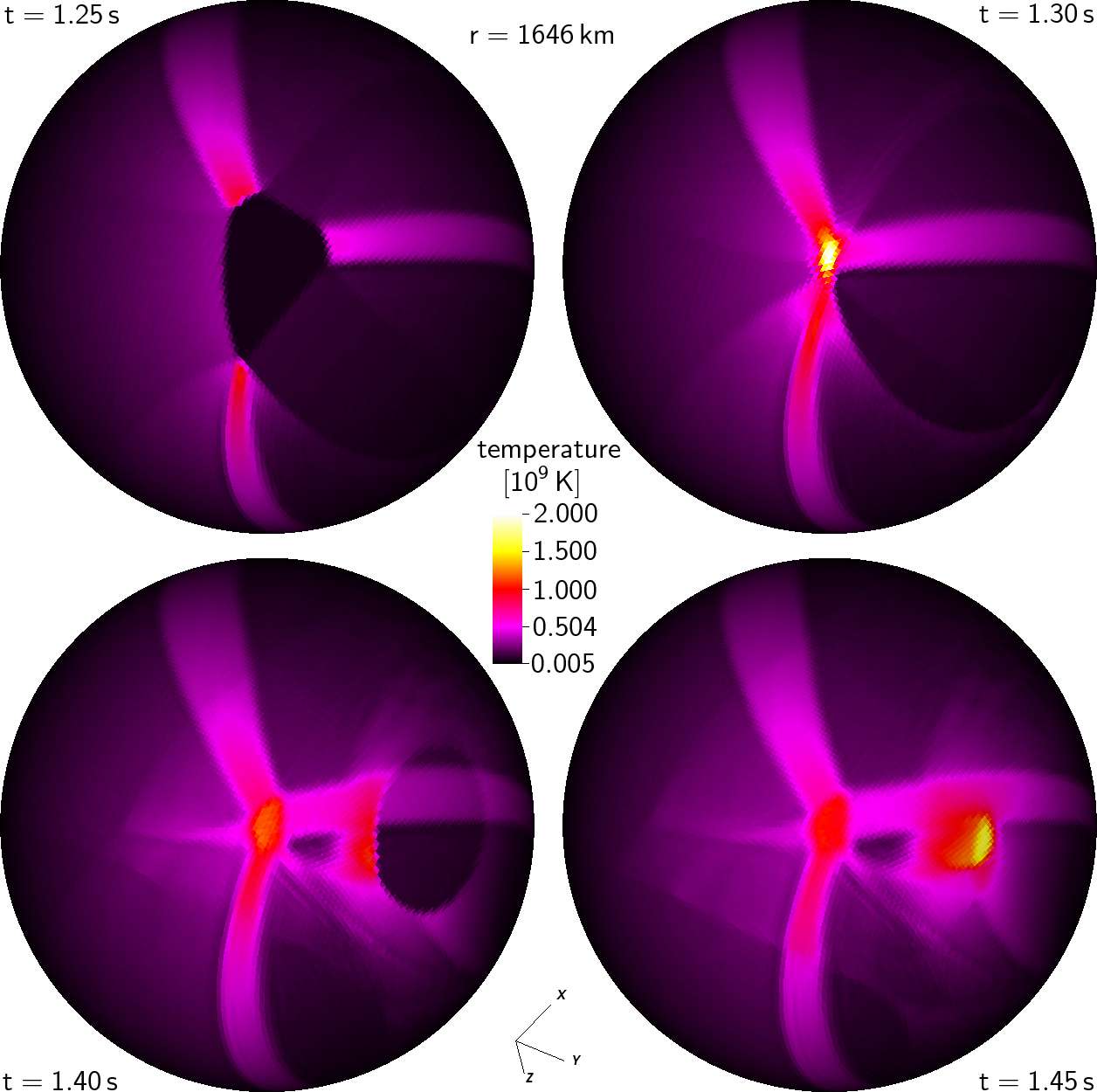}
\caption{Snapshots of the temperature in a simulation with three detonators
(run~\ref{sim:h0529x}) on a concentric sphere inside the core which contains
the hot spot from the three converging waves (upper right panel) as well as the
hot spot from the wave induced by the first detonator alone (lower right
panel).}
\label{fig:tsphere}
\end{figure}

To study the sliding helium detonation and the internal compressional wave
which potentially produces the seed for a core detonation, we suppress nuclear
burning in zones with more than $10\%$ \el{C}{12}. This is to prevent spurious
carbon burning at the core-shell boundary due to the mixing of hot helium with
cold carbon in a zone, and to prevent core detonation before the waves in the
core have fully converged and produced the highest possible temperatures and
densities (which we want to determine).  For the sake of comparability, we here
focus on the $(0.801+0.143)\Msun$ model~\ref{mod:p8b1a} for the most part. The
detonation of WDs with $1.00\Msun$ cores and lightweight helium shells is
discussed in \S\ref{sec:lighthesh}.

\subsubsection{One Detonator}

We here consider the case of a single, spherical detonation spot setting off a
helium detonation where the hottest layer (at $r_8=4.27$) intersects with the
positive $z$-axis in our coordinate system.  This axisymmetric problem was
investigated with different resolutions in three 2D runs (\ref{sim:p0403x},
\ref{sim:p0229x}, and \ref{sim:p0406x}), and one 3D run (\ref{sim:p0324x}) in a
quarter-star sized domain and mirror symmetries at the inner boundaries.

The helium detonation wave wraps around the star (see Figure~\ref{fig:hotspot}a)
and reaches the antipode of the detonator after about $1.2\unit{s}$.  For
comparison, a sound wave in the helium shell would need about $5.6\unit{s}$ to
reach the other side of the star. This corresponds to a front velocity of $\sim
1.1\times10^9\cms$.  The polar (latitudinal) velocity of the ash directly
behind the front and the sound speed in the ash both are roughly
$6.9\times10^8\cms$. Their sum is larger than the actual front velocity,
consistent with incomplete burning rather than a Chapman--Jouguet detonation
\citep[cf.][]{2012Sim}. The \el{Ni}{56} fraction far ($\simm 1000\unit{km}$) behind
the detonation front is about 70\%.

The sliding detonation induces an internal compressional wave which converges
off-center inside the core on the $z$-axis (in general: the axis of symmetry in
the problem).  The average velocity with which the perturbation propagates in
negative $z$-direction from the helium ignition point to the final convergence
point is $\sim 4.0\times10^8\cms$, a little larger than the mean sound speed
inside the core ($3.4\times10^8\cms$).

The temperatures and densities in some zones of run~\ref{sim:p0403x} are high
enough to potentially\footnote{Whether a spontaneous detonation initiates
depends not only on the density, the peak temperature and the size of the
region (or, equivalently, its mass), but also on the shape of the temperature
profile (as emphasized by \citealt{2009Seitenzahl}).  The conditions in the
hottest zone are listed in Table~\ref{tab:hspots}.  Since the grid zones in our
simulations are usually larger than the critical masses for detonation, it is
not possible to predict with certainty whether the conditions in one or a few
zones lead to detonation.} trigger a core detonation where the internal wave
finally converges \citep{1997Niemeyer,2007Roepke,2009Seitenzahl}.  However, it
is likely that the core ignites earlier, at larger distance from the center and
close to the core-shell interface on the $z$-axis, where the internal wave
front converges in the polar direction (but not in radial direction), and the
converging helium detonation drives a hot, thin inflow of \el{Ni}{56} in positive
$z$-direction.  As a numerical experiment, we restarted run~\ref{sim:p0403x}
without suppressing the burning of carbon-rich zones at times $t=1.2\unit{s}$
(when the helium detonation converges), $1.3\unit{s}$, $1.4\unit{s}$, and
$1.5\unit{s}$ (when the internal wave converges).  In all of these cases, a
detonation wave in the core forms immediately.

The values of the density and temperature in the hot spot are dependent on the
spatial resolution $\Delta x$ of the grid as well as on the temporal separation
$\Delta t$ of the considered snapshots (for practical reasons, we do not edit
and analyze the results at every time step), cf. the values of temperature and
density in the hot spots of runs~\ref{sim:p0403x}, \ref{sim:p0229x},
\ref{sim:p0406x} and~\ref{sim:p0324x} listed in Table~\ref{tab:hspots}.  However,
all results unequivocally indicate that the core would detonate.  We expect
that a higher resolution would only increase the likelihood for detonation.

\subsubsection{Opposite Detonators}

As a limiting case of multi-point ignition, we at first consider two detonators
at opposite points in the helium shell (run~\ref{sim:p0422x}).  While it is
unlikely for such detonators to be synchronous, this poses a computationally
cheap 2D problem which can be regarded as a numerical experiment.  The internal
waves in this case collide on the $z=0$ plane (the detonators being centered on
the $z$-axis at $z_8=\pm 4.27$), beginning at the outer edge of the core at
$t=0.60\unit{s}$ and reaching the center at $t=1.05\unit{s}$.  The final wave
collision in the central part of the star takes place on a $\simm1400\unit{km}$
wide disk without generating a discernible hot spot.  However, temperatures
$\gtrsim 10^9\unit{K}$ occur on the collision plane at large radii ($r_8 \sim
1.8\ldots3.2$).  While the material is not compressed there, we observe a
radially inward moving core detonation in a run where the burning of carbon is
allowed for $t\geq0.65\unit{s}$.  If carbon burning is suppressed, the original
core waves are reflected at the $z=0$ plane. The reflected waves collide on the
$z$-axis (symmetrically on both sides of the $z=0$ plane), creating dense hot
spots at $z_8=\pm 1.45$ that would very likely have ignited the core if
ignition had failed earlier.

When one of the detonators (in our case the one on the negative side of the
$z$-axis) is delayed by $0.30\unit{s}$, the two helium detonation fronts
collide at $t=0.75\unit{s}$ at an angle of $112\degree$ with respect to the
positive $z$-axis, see Figure~\ref{fig:p0425}. As in the synchronous case, the
collision of the primary internal waves, which here happens on a slightly
curved disk at $z \approx -600\unit{km}$, does not generate a hot spot.
However, a wave reflected toward the negative $z$-axis produces a hot spot
that is sufficiently dense to trigger core detonation.

\subsubsection{Two Synchronous Detonators}

Before discussing 3D cases of two-point ignition in the helium, it is worth
noting that a single, small aspherical detonator is virtually equivalent to a
spherical one, regardless of how strongly deformed it is (as long as it sets
off a detonation). As the detonation expands with constant velocity in every
direction, the original asphericity is quickly lost.  To assess the effects of
non-axisymmetry on the focusing of the internal waves, we therefore consider
cases with two detonators at varying but significant separation from one
another.  For two synchronous detonators, it suffices to do a 3D simulation of
a quarter of the star, with mirror symmetry at the inner boundaries.

In cases with two moderately separated, synchronous detonators (we consider
angular separations of $36\degree$, $54\degree$ and $90\degree$), the initially
separate helium detonation fronts collide halfway between the two spots on the
geodesic connecting them on the core surface, in our case the positive
$z$-axis.  While the detonations merge, the combined front proceeds toward the
antipode of this collision point.  The front initially has a head start on the
plane that connects the two spots with the center of the star (the $x=0$ plane
in our simulations), but the deficit becomes smaller as the detonations
continue to merge\footnote{For comparison, consider two merging detonations in
a plane, starting from the $x$-axis at $\pm a$. With $r$ being the radius of an
individual detonation, the height of the combined detonation on the $y$-axis is
$\sqrt{r^2 - a^2}$, asymptotically approaching $r$ (and not $r-a$) at large
radii.  That is, the two initially separate detonations become
indistinguishable when their size is much larger than the initial separation.}.
As it approaches the antipode, the detonation front has the shape of a pointed
ellipse.  This shape can easily be understood considering the combined
detonation as a superposition of two detonations: the pointed ellipse is the
intersection between two circles (or, more precisely, lines of constant
latitude, if the ignition points are regarded as poles) whose centers are the
respective antipodes of the ignition points.  In a simulation, the detonation
finally converges when the width of the pointed ellipse approaches the size of
a grid zone.  Theoretically, the ratio of the length of the ellipse compared to
its width increases indefinitely, i.e., the wave always converges in a line
rather than a point and the difference between having one and two detonators
should become larger with increasing grid resolution or diffusion length.

In cases with $36\degree$ and $54\degree$ detonator separation, the region in
which the internal waves converge becomes sufficiently hot and dense to trigger
a core detonation, see Table~\ref{tab:hspots} and Figure~\ref{fig:hotspot}b.  We
explicitly confirmed this for run~\ref{sim:p0322x} by switching on carbon
burning for $t>1.40\unit{s}$.  When two synchronous detonators are $90\degree$
apart (run~\ref{sim:h0430x}), there is no hot spot at the site where the two
primary waves converge at the $z$-axis (in general: the intersection of the two
planes of symmetry in this problem).  However, the waves induced by the two
detonators pass through one another and form two hot spots at considerable
distance ($660\unit{km}$) from the $z$-axis, see Figure~\ref{fig:hotspot}c.

\subsubsection{Two Asynchronous Detonators}

We next consider a case of two asynchronous detonators with an angular
separation of $54\degree$, run~\ref{sim:h0427x}.  Note that this scenario is
less symmetric than the case of two synchronous detonators, there being only
one mirror symmetry in the problem, across the plane which intersects both
detonators.  We therefore use a 3D domain that comprises half of the star and
assume mirror symmetry at the inner boundary.  One of the detonators---the one
on the $+y$ side in our simulations---is set off $0.15\unit{s}$ (roughly half
the time needed for the first detonation to reach the second detonator) after
the first one\footnote{We implemented this in the code by suppressing the
time-step update of the zones comprising the detonation spot for said time.}.
Shortly before converging, the helium detonation front assumes the shape
defined by the intersection of two circles with different radii, see
Figure~\ref{fig:h0427threed}.  The same shape can be seen in the internal wave
front on concentric spherical surfaces.  Just like in the synchronous case
(run~\ref{sim:h0407x}), a dense hot spot forms when the internal wave fronts
converge, see Figure~\ref{fig:h0427x} and the rightmost snapshot in
Figure~\ref{fig:h0427threed}.

With $120\degree$ separation and a delay of $0.30\unit{s}$,
run~\ref{sim:j0510x}, there is no hot spot where the internal waves from the
two detonators converge. As in the case with synchronous detonators at
$90\degree$ discussed above, the waves from the two detonators pass through one
another.  The wave from the first detonator is the first to converge and
trigger a core detonation.

\subsubsection{Three Asynchronous Detonators}

Unlike the above cases, a setup with three or more detonators is in general
free of symmetries and requires a 3D simulation of the full star.  In
run~\ref{sim:h0529x}, we again consider two detonators (A,B) with a separation
of $120\degree$, one of which (B) is delayed by $0.30\unit{s}$, exactly as in
run~\ref{sim:j0510x}. In addition, a third detonator C, delayed by
$0.20\unit{s}$, is placed $90\degree$ from detonator A and $100\degree$ from
detonator B on the positive $x$ side.  In this case, when the helium detonation
converges toward the last patch of unburnt helium, it roughly has a triangular
shape. The same shape is present in the internal wave on concentric spherical
surfaces.  There is a distinct hot spot when the internal waves converge.
Shortly after, a secondary hot spot arises as the wave induced by the first
detonator converges again with itself.  Figure~\ref{fig:tsphere} shows snapshots
of both hot spots (right hand panels) and the converging wave fronts from which
they originate (left hand panels).

\subsubsection{Light Helium Shells}
\label{sec:lighthesh}

\begin{figure}[t]
\center
\includegraphics[width=.8\linewidth]{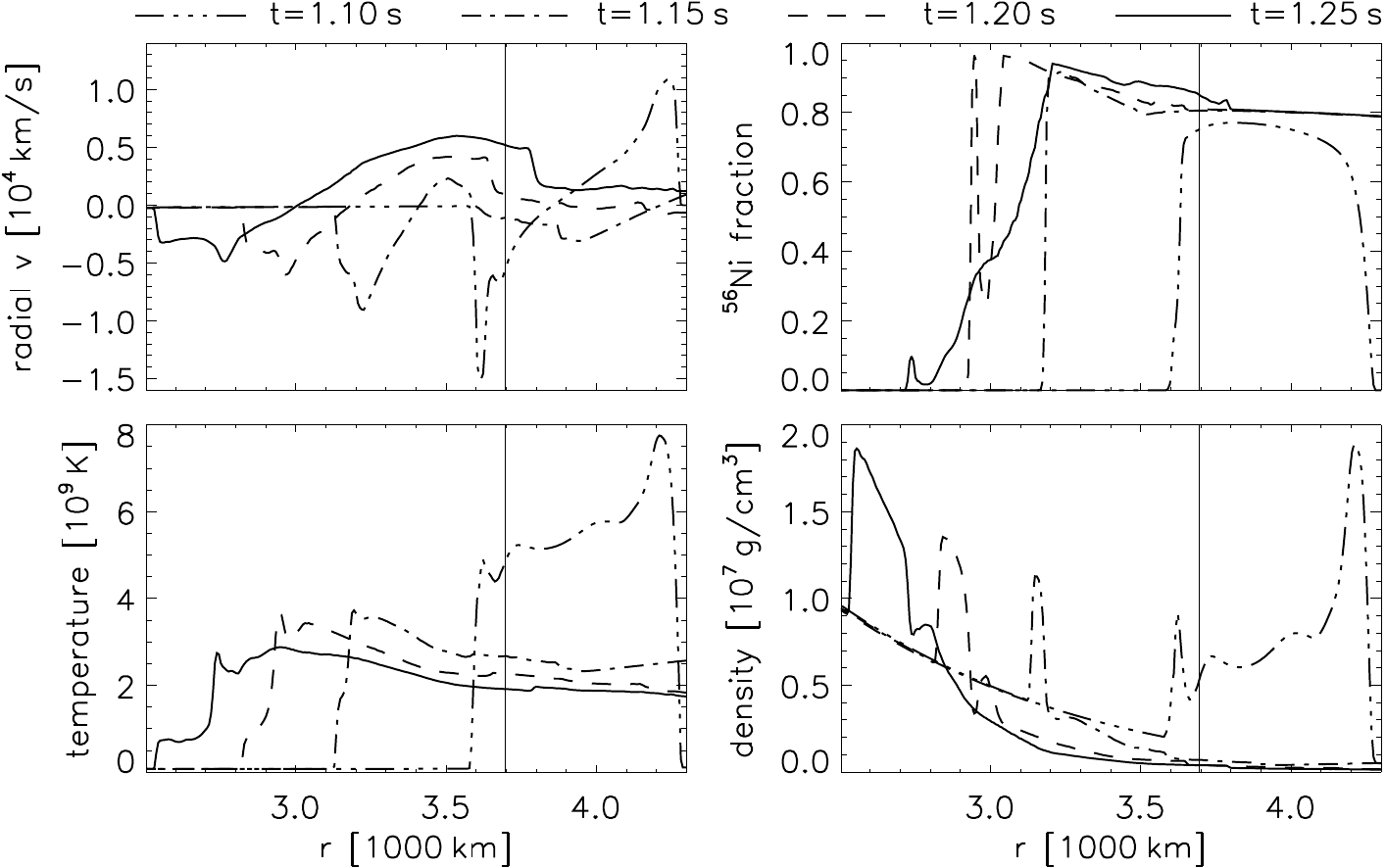}\\[\medskipamount]
\includegraphics[width=.8\linewidth]{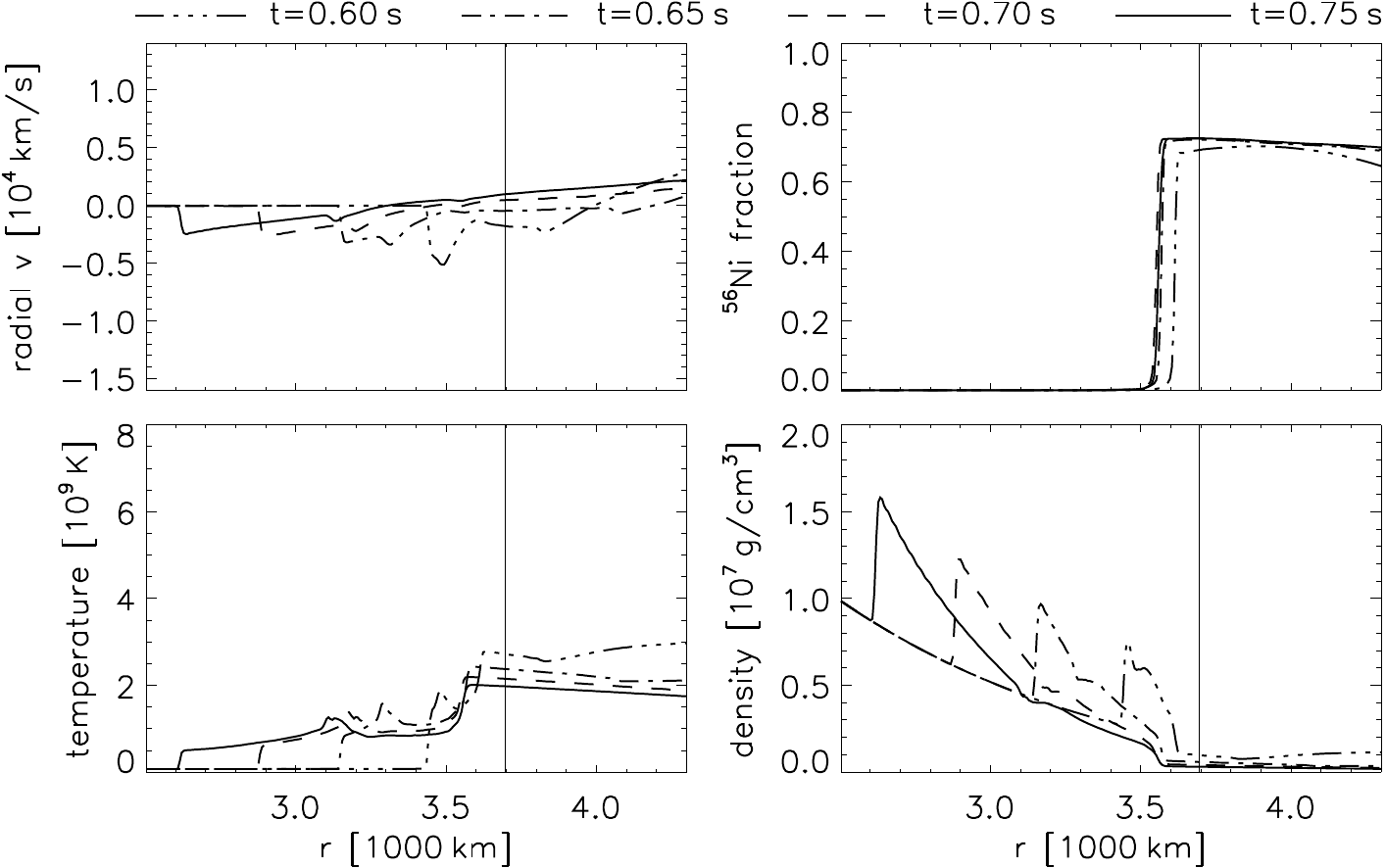}
\caption{Conditions near the core boundary as a function of distance from the
center. Top panel: on the line (negative $z$-axis) where the helium detonation
front converges in a run with one detonator (run~\ref{sim:p0514x}).  Bottom
panel: on the helium collision plane ($z=0$) in a run with two opposite
synchronous detonators (run~\ref{sim:p0515x}).  The vertical lines mark the
initial radius of the core-shell interface.  The plots represent calculations
where carbon burning is suppressed; otherwise the core ignites at the earliest
time (dot-dot-dot-dashed line) in both cases.}
\label{fig:jets}
\end{figure}

\begin{figure}[t]
\includegraphics[width=\linewidth]{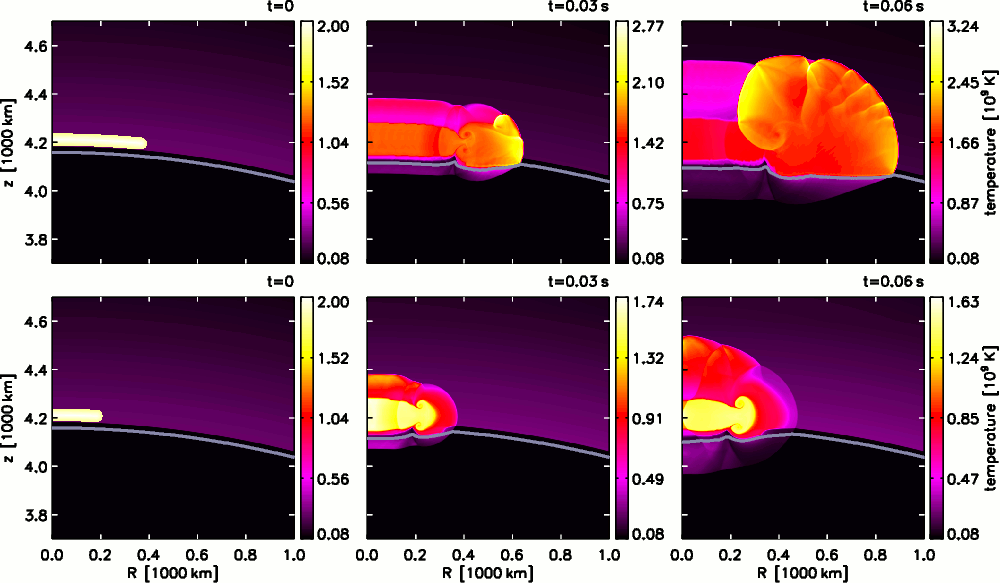}
\caption{Successful (top) and unsuccessful (bottom) initiation of a detonation
in the helium shell of model~\ref{mod:1pi1a} with a detonator in the form of a
spherical cap. The cases shown are the last two in Table~\ref{tab:caps}.  The
color bar ranges from the minimum to the maximum temperature in each plot (note
the decline of the maximum in case of failure). The gray contour line, at 49\%
\el{C}{12}, marks the core-shell interface.}
\label{fig:caps}
\end{figure}

\begin{deluxetable}{ccccc}
\tablecaption{Detonations started in a spherical cap}
\tablehead{
  altitude\tablenotemark{a} & arc length & \multirow{2}{*}{rim density} & rim velocity & \multirow{2}{*}{detonates} \\
  {[km]}                      & [km]       &                              & [$10^8\cms$]  &
}
\startdata
0 & 141 & 2 $\times$ ambient & 8 & yes \\
0 & 96  & 2 $\times$ ambient  & 8 & no \\
0 & 233 & ambient & 8 & yes \\
0 & 141 & ambient & 8 & yes\tablenotemark{b} \\
0 & 96  & ambient & 8 & no \\
0 & 415 & ambient & 1 & yes \\
0 & 233 & ambient & 1 & no \\
0 & 415 & ambient & 0.5 & no \\
0 & 781 & ambient & 0.5 & yes\tablenotemark{b} \\
25 & 785 & ambient & 1 & yes \\
25 & 418 & ambient & 1 & no
\enddata
\tablenotetext{a}{Radial distance from the hottest helium layer to the center of the cap}
\tablenotetext{b}{Detonation moves only inward at first and is then reflected back}
\label{tab:caps}
\end{deluxetable}

We here consider axisymmetric detonations (one detonator or two opposite
detonators) for progenitor models other than \ref{mod:p8b1a}.  A detonation in
a $0.078\Msun$ helium shell on top of a solar mass core (model~\ref{mod:1pj1a},
run~\ref{sim:p0514x}) is set off by a $25\unit{km}$ spherical detonator
centered on the hottest helium layer.  The sliding helium detonation yields a
distinct hot spot where the internal wave converges at $t \approx
1.40\unit{s}$.  With carbon burning turned on from the time when the helium
detonation converges ($t \approx 1.10\unit{s}$), the core immediately detonates
near its surface.  This detonation is triggered by a thin, hot radial inflow of
\el{Ni}{56}. As shown in the upper panel of Figure~\ref{fig:jets}, temperatures
of $2\text{--}4\times10^9\unit{K}$ and densities on the order of $10^7\gcc$ are
generated along this jet.  In a simulation with two synchronous, opposite
detonators, run~\ref{sim:p0515x}, temperatures $> 10^9\unit{K}$ are generated
neither when the wave fronts converge at the center, nor along the $z$-axis
when the reflected waves collide (in contrast with model~\ref{mod:p8b1a},
run~\ref{sim:p0422x}).  The only region hot enough to light the core in this
case is located at the equatorial plane near the core's surface, where the
collision of the internal waves begins.  Although there is no hot inflow
comparable to the case with one detonator (note that a radial inflow in this
case is not a jet), see lower panel in Figure~\ref{fig:jets}, the core there
ignites in the simulation.

It is difficult to set off a detonation in a $0.045\Msun$ helium shell
sorrounding a one solar mass core (model~\ref{mod:1pi1a}).  We did not succeed
with a spherical detonator centered on the hottest helium layer, whose
density is $7.17\times10^5\gcc$.  A detonation does not start even
with a (perhaps unrealistically) strong initial detonator: a central
temperature of $3\times10^9\unit{K}$, linearly decreasing to
$1.5\times10^9\unit{K}$ at the outer edge, a density three times that
of the ambient medium and a radius of $25\unit{km}$ (ending
$1\unit{km}$ above the core-shell interface).  Large shock velocities
(we tried $1.6\times10^9\cms$ instead of our usual $8\times10^8\cms$)
at the outer edge of the detonation spot did not help either.

A detonation in the $0.045\Msun$ shell of model~\ref{mod:1pi1a} can be set off,
however, by a large spherical detonator of radius $50\unit{km}$ (which is about
twice the distance between the core-shell interface and the hottest helium
layer) centered $25\unit{km}$ above the hottest helium layer. The resulting
detonation front is more feeble than in all cases with heavier helium shells,
blowing out at too coarse a resolution ($\mathord{\lesssim}2\unit{km}$ works,
$6.5\unit{km}$ does not), and requiring the use of AMR in our simulated helium
shell. But once started, the detonation propagates around the star and
converges after $\approx 1.40\unit{s}$ (run~\ref{sim:p0627x}). The
corresponding front velocity is $\sim 9.4\times 10^8 \cms \approx$ 4.3 times
the sound speed in the hottest helium layer.  When the internal wave induced by
the helium detonation converges $0.40\unit{s}$ later, it forms a distinct hot
spot.  Without suppression of the carbon burning, the core detonates
immediately after the convergence of the helium detonation, starting at the
$z$-axis near the core-shell boundary.  In a simulation with two synchronous,
opposite detonators, run~\ref{sim:p0625x}, no hot spot is formed in the core.
With carbon burning turned on, the core does not detonate in the simulation.

\subsection{Ignition in an Extended Region}

As stated in the above paragraph, a spherical detonator in
model~\ref{mod:1pi1a} needs to be fairly large to set off a helium detonation.
This, however, is not compatible with a runaway starting at the hottest layer
in the helium shell.  As an alternative, we here consider the possibility of a
thermonuclear runaway starting in a (curved) sheet on an equipotential, instead
of a single point.  The size of convective cells in the helium shell is
expected to be on the order of the pressure scale height, which is
$366\unit{km}$ at the hottest helium layer in model~\ref{mod:1pi1a}.  The size
of the region where the runaway starts could be a substantial fraction of this,
but is probably not larger.  As a toy model, we consider detonation spots in
the form of spherical caps with a thickness of $50\unit{km}$. Selected runs are
listed in Table~\ref{tab:caps} and examples for a successful and an
unsuccessful detonation are shown in Figure~\ref{fig:caps}.  As in most of our
detonators, the temperature decreases linearly from $2\times10^9\unit{K}$ at
the center to $1.8\times10^9\unit{K}$ at the outer edge (on all sides of the
cap). The outer rims of the caps are half tori, imposed with a velocity that
increases linearly from the torus center to the outer edge. An arc length of
$50\unit{km}$ constitutes a limiting case in which the cap degenerates into a
sphere with the same properties as our spherical detonators.

The cap detonator is more powerful than a spherical one with the same radial
(with respect to the center of the star) extent.  Detonations can be initiated
at ambient density with subsonic starter velocities. Smaller sizes require
harsher starting conditions.  Below a size of $\simm 100\unit{km}$ (roughly 1/3
pressure scale height), the initiation of a detonation appears impossible.  For
the given density of $\simm7\times10^5\gcc$ near the temperature maximum of
model~\ref{mod:1pi1a}, this limit is in approximate agreement with the recent
findings of \citet{2013Holcomb}, who determined the critical sizes of helium
detonation for a wide range of densities in a series of 1D simulations.

\subsection{Complete Detonations}
\label{sec:complete}

\begin{deluxetable}{cccc}
\tablecaption{Nucleosynthesis yields in the helium shell [$\Msun$]}
\tablehead{
         & Model~\ref{mod:p8b1a} & Model~\ref{mod:1pj1a} & Model~\ref{mod:1pi1a}
}
\startdata
    \el{C}{12}  & $8.00\times10^{-4}$ & $5.24\times10^{-4}$ & $7.02\times10^{-4}$ \\
    \el{N}{14}  & $1.11\times10^{-10}$ & $2.83\times10^{-14}$ & $4.24\times10^{-13}$ \\
    \el{O}{16}  & $3.19\times10^{-4}$ & $1.93\times10^{-4}$ & $2.46\times10^{-5}$ \\
    \el{Ne}{20} & $4.78\times10^{-5}$ & $2.72\times10^{-5}$ & $1.95\times10^{-5}$ \\
    \el{Mg}{24} & $9.40\times10^{-5}$ & $5.30\times10^{-5}$ & $3.00\times10^{-5}$ \\
    \el{Si}{28} & $4.46\times10^{-4}$ & $2.65\times10^{-4}$ & $2.03\times10^{-4}$ \\
    \el{S}{32}  & $2.88\times10^{-4}$ & $1.26\times10^{-3}$ & $2.17\times10^{-4}$ \\
    \el{Ar}{36} & $4.55\times10^{-4}$ & $2.65\times10^{-4}$ & $4.62\times10^{-3}$ \\
    \el{Ca}{40} & $2.15\times10^{-3}$ & $1.26\times10^{-3}$ & $2.17\times10^{-3}$ \\
    \el{Ti}{44} & $4.14\times10^{-3}$ & $2.33\times10^{-3}$ & $4.29\times10^{-3}$ \\
    \el{Cr}{48} & $4.24\times10^{-3}$ & $2.36\times10^{-3}$ & $4.39\times10^{-3}$ \\
    \el{Fe}{52} & $7.81\times10^{-3}$ & $4.34\times10^{-3}$ & $5.69\times10^{-3}$ \\
    \el{Fe}{54} & $1.80\times10^{-5}$ & $2.57\times10^{-6}$ & $5.16\times10^{-7}$ \\
    \el{Ni}{56} & $7.05\times10^{-2}$ & $3.71\times10^{-2}$ & $1.95\times10^{-3}$
\enddata
\label{tab:heyields}
\end{deluxetable}

\begin{figure}[t]
\centering
Model~\ref{mod:1pj1a} \, $(1.001+0.078)\Msun$\\
\includegraphics[width=.7\linewidth]{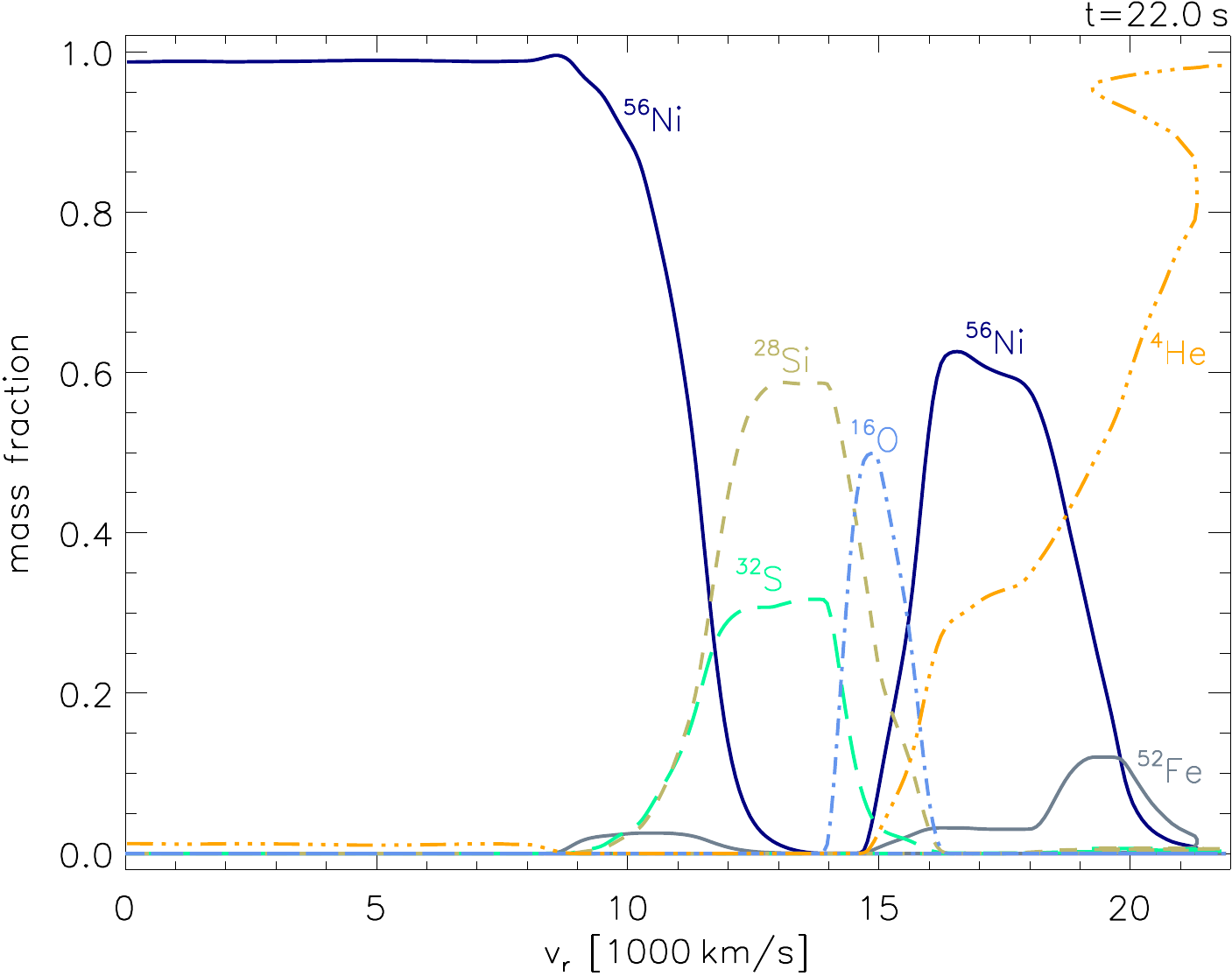}\\[\medskipamount]
Model~\ref{mod:1pi1a} \, $(1.002+0.045)\Msun$\\
\includegraphics[width=.7\linewidth]{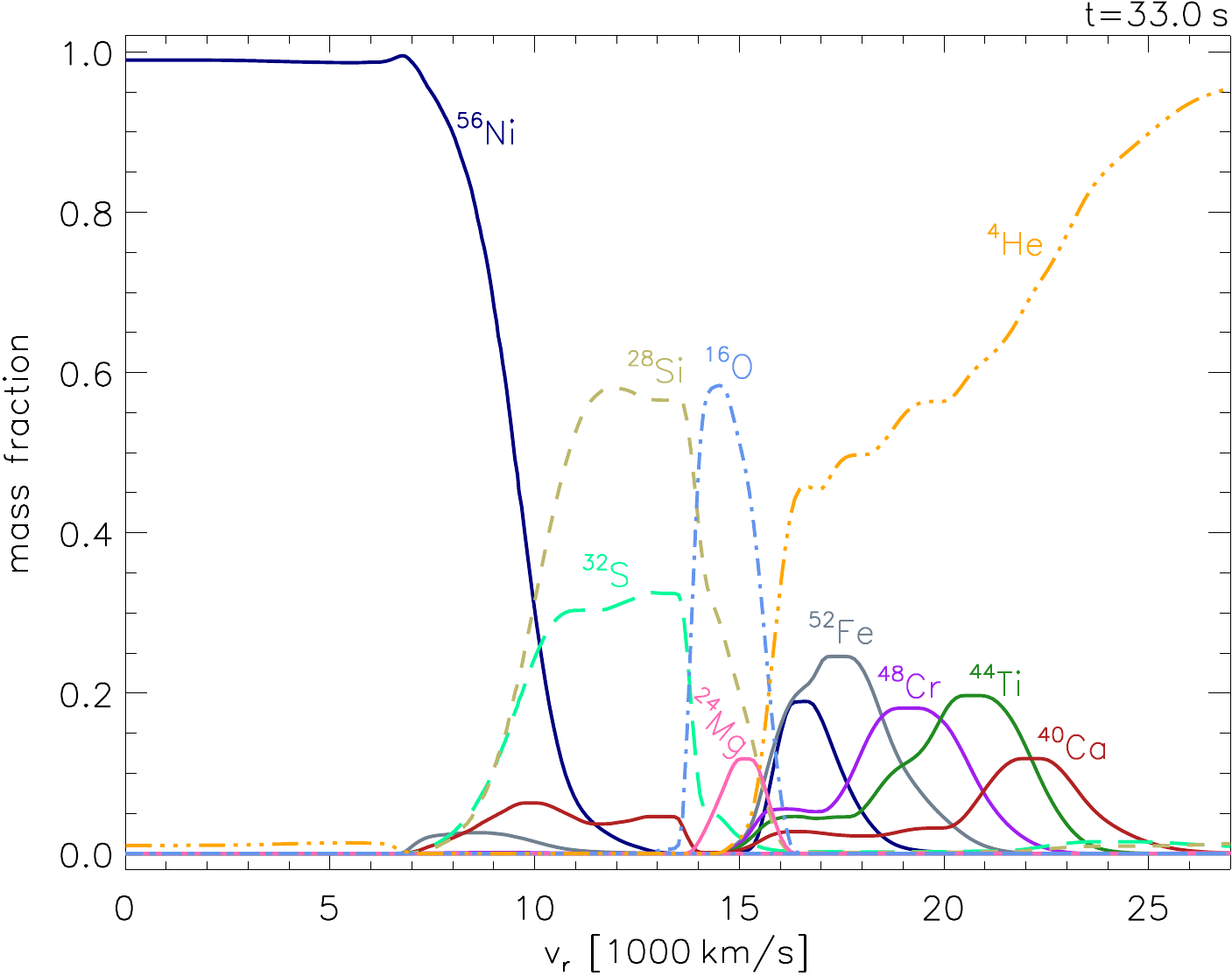}
\caption{Mass fractions of different isotopes along the equatorial plane
($z=0$) in the ejecta of models~\ref{mod:1pj1a} (top panel) and~\ref{mod:1pi1a}
(bottom panel) after complete shell and core detonations, as functions of
radial velocity (which is approximately proportional to radius). In both
panels, only isotopes with maximum fractions $> 10\%$ are plotted,
respectively.}
\label{fig:els}
\end{figure}

Table~\ref{tab:heyields} lists the yields from helium shell detonations in
different models, measured at the time when the helium detonation converges
opposite the detonator (we only consider single detonators here, not expecting
the yields from detonations started at multiple points to be much different).
The lightweight helium shell in model~\ref{mod:1pi1a} produces little
\el{Ni}{56}, only about 5\% of what the heavier shell in model~\ref{mod:1pj1a}
produces, but the yields of isotopes with mass numbers between \el{Ar}{36} and
\el{Fe}{52} is in total about twice as high as in model~\ref{mod:1pj1a}.

We followed the detonation of the core and the expansion of the ejecta to
scales of $10^6\unit{km}$ (about 200 times the size of the initial WD), at
which time the expansion is homologous.  The total nickel yields in
models~\ref{mod:p8b1a}, \ref{mod:1pi1a} and~\ref{mod:1pj1a} is $0.38\Msun$,
$0.52\Msun$ and $0.66\Msun$, respectively.  Mass fractions of selected isotopes
in models~\ref{mod:1pj1a} and~\ref{mod:1pi1a} are plotted in
Figure~\ref{fig:els}. The inner parts of the core turn into almost pure
\el{Ni}{56} in both cases, whereas the outer parts yield mainly \el{Si}{28} and
\el{S}{32}.  The lighter helium shell produces large fractions of \el{Ca}{40},
\el{Ti}{44}, \el{Cr}{48} and \el{Fe}{52} instead of \el{Ni}{56}, which is the
dominant product of the heavier shell.

\section{Summary and Discussion}
\label{sec:discussion}

We studied the double detonation scenario for Type Ia supernovae by means of 2D
and 3D simulations. Starting from the results of 1D stellar evolution
simulations, we find that the helium detonation wave is halted at the
core-shell interface if the detonation starts in the hottest layer, which in
general is located at some altitude above the interface. If the detonation
commences at higher altitude, the detonation may transcend into the core.  For
direct drive to be successful, the helium detonation must involve a critical
mass which is sensitive to the density at the core-shell interface.  A layer of
mixed core-shell material does not facilitate the direct drive.  In marginal
cases, the detonation transcends the core-shell interface but then fragments,
opening up the possibility of incomplete core detonations.

We confirm that the sliding helium detonation induces a mildly supersonic
compressional wave inside the core which, if the detonation is set off at a
single point in the helium shell, converges to produce an off-center spot that
is sufficiently hot and dense to light the core as well.  We tested the
robustness of this model for non-symmetric initial conditions, considering a
series of scenarios in which the helium detonation is started at two points and
one scenario in which it is started at three points.  We find that the
secondary core detonation is quite robust, despite the lack of symmetry. In
cases where detonators are widely separated ($\mathord{\gtrsim}90\degree$), the
hot spot is produced by reflected waves or waves that have passed through other
waves before converging with themselves, rather than the converging primary
waves.  Only in one extreme case---two antipolar, synchronous detonators in the
helium shell of a $(1.002+0.045)\Msun$ WD---we find no hot spot and no core
ignition.

The generation of a detonation-inducing hot spot is robust even in a completely
non-symmetric setup with three asynchronous detonators.  Geometric arguments
suggest that it is even more robust than a setup with two detonators, where the
helium detonation and the core waves converge in an elongated region. As the
elongation increases with the smallest possible separation of the detonation
fronts, the difference of having one, two or three detonators might be larger
at higher resolution.  Although we have not run cases with more than three
detonators, we expect that the core ignition would not be thwarted by
additional detonators.  A case with numerous detonators may be similar to a
detonation in a spherical shell.

After the helium detonation has converged, the internal waves converge in polar
(latitudinal) direction at ever decreasing radii, beginning right beneath the
core-shell interface. In addition, the collision of the helium detonation front
can drive a jet of hot ash into the core.  In any case, this region has high
ignition potential. That is, the core may ignite at large radii before the
``proper'' hot spot forms deeper inside as the wave fronts converge also in the
radial direction.  The complete convergence of internal waves may only happen
in cores that are difficult to ignite.

Our simulations suggest that it is very hard, perhaps impossible under
realistic assumptions, to initiate a helium detonation at densities
$\lesssim10^6\gcc$ (as observed in plane-parallel geometry by
\citealt{2012Townsley}).  On the other hand, the yields from WDs with thin
helium shells are most compatible with observed Type Ia supernovae.
Detonations in the helium are more easily started in an extended sheet rather
than a point.  Our simulations indicate that the size of such a sheet would
have to be at least on the order of the size of a convective cell in the helium
shell of a quasi-stationary WD. Whether detonation seeds of this kind are
realistic can ultimately be answered only by suitable 3D convection
simulations.  Alternatively, a sufficiently large region may reach the
conditions for helium detonations in the context of double-degenerate mergers
(which are not directly considered in this work), by the interaction of an
accretion stream with previously accreted helium \citep{2010Guillochon} or
during the merger itself \citep{2012Raskin}.

\begin{acknowledgments}
This research has been supported by the DOE SciDAC Program under contract
DE-FC02-06ER41438; the National Science Foundation (AST 0909129) and the NASA
Theory Program (NNX09AK36G). We acknowledge useful discussions with Haitao Ma
concerning the implementation of detonation physics in CASTRO. We also thank
John Bell and Ann Almgren for their major roles in developing the CASTRO code.
R. Moll acknowledges support by the Alexander von Humboldt Foundation through
the Feodor Lynen Research Fellowship program. This research used resources of
the National Energy Research Scientific Computing Center, which is supported by
the Office of Science of the U.S. Department of Energy under Contract No.
DE-AC02-05CH11231.  This research used resources of the Oak Ridge Leadership
Computing Facility at the Oak Ridge National Laboratory, which is supported by
the Office of Science of the U.S. Department of Energy under Contract No.
DE-AC05-00OR22725.
\end{acknowledgments}

\bibliography{ref}

\end{document}